\documentclass[twocolumn]{aastex7}

\usepackage{amsmath}
\usepackage{graphicx}
\usepackage{textcomp}
\usepackage{gensymb}
\usepackage{natbib}
\usepackage{longtable}
\usepackage{multirow}
\usepackage{float}
\usepackage{enumitem}

\begin{document}
\def\teff{$T_{\rm eff}$}
\def\cs{$\chi^{2}$}
\def\rsun{$R_{\odot}$}
\def\msun{$M_{\odot}$}
\def\rstar{$R_{\star}$}
\def\rearth{$R_{\earth}$}
\def\av{$A_{V}$}
\def\emcee{\texttt{emcee}}
\def\kepler{\textit{Kepler}}

\graphicspath{{figures/}}
\shorttitle{Occurrence Rates for Planets in Binaries}
\shortauthors{Sullivan et al.}

\title{The First Occurrence Rate Estimates for Exoplanets in Small-Separation Binary Star Systems: Planet Occurrence is Suppressed in Binary Stars}

\author[0000-0001-6873-8501]{Kendall Sullivan}
\affil{Department of Astronomy and Astrophysics, University of California Santa Cruz, Santa Cruz CA 95064, USA}
\email{ksulliv4@ucsc.edu}

\author[0000-0002-1092-2995]{Anne Dattilo}
\affiliation{Department of Astronomy \& Astrophysics, 525 Davey Laboratory, 251 Pollock Road, Penn State, University Park, PA, 16802, USA}
\affiliation{Center for Exoplanets and Habitable Worlds, Penn State University, 525 Davey Laboratory, 251 Pollock Road, University Park, PA, 16802, USA}
\email{adattilo@psu.edu}

\author[0000-0002-7030-9519]{Natalie M. Batalha}
\affil{Department of Astronomy and Astrophysics, University of California Santa Cruz, Santa Cruz CA 95064, USA}
\email{natalie.batalha@ucsc.edu}

\correspondingauthor{Kendall Sullivan}
\email{ksulliv4@ucsc.edu}

\begin{abstract}
Exoplanet occurrence rates facilitate comparisons between observations of planets and theoretical models of planet formation. Despite their deductive power, exoplanet occurrence rates for half the stars in the sky are missing because occurrence rate studies systematically exclude binary star systems. We assembled a large sample of high-likelihood binaries from the Kepler mission to calculate occurrence rates for circumstellar (S-type) planets in small-separation binary star systems ($\lesssim 100$ au) for the first time. For a sample of high-likelihood small-separation binaries, we found binaries to host 58\% fewer planets per system than single stars to 11.4$\sigma$ significance within 1-4\,$R_{\oplus}$ and 1-50\,d, and 50\% fewer planets compared to single stars when integrating over the full parameter space of 1-10\,$R_{\oplus}$ and 1-100\,d to 3.8$\sigma$ significance.. We found no evidence for a radius valley or radius cliff, instead detecting a smooth decline in planet occurrence with increasing planetary radius. The difference between the single-star planet radius distribution and the binary-star planet radius distribution is 4.3$\sigma$ significant from a Kolomogorov-Smirnov test. These results suggest significantly different planet formation and survival outcomes in binaries compared to single stars, and support other studies that have measured a deficit of observed planets in binary star systems.
\end{abstract}

\keywords{}

\section{Introduction}
As the number of known exoplanets has grown, two regimes of exoplanet science have become common: bespoke, in-depth studies of single objects or small samples; and studies of the properties of large populations of exoplanets. The second of these efforts is known as exoplanet demographics. Studies of demographics offer the opportunity to understand what the typical outcomes of planet formation are, and large planet samples let us explore a wide parameter space as well as compare observations to theories of planet formation. 

Exoplanet demographics have revealed many important features of the planet population, all of which can be linked to important planetary physics. For example, the Neptune desert \citep[e.g.,][]{Mazeh2016}, the radius valley \citep[e.g.,][]{Fulton2017}, and the radius cliff \citep[e.g.,][]{Dattilo2024} are all important signatures of formation and evolutionary differences between different exoplanet populations. 

However, demographic features seen in observed populations do not necessarily correlate to the underlying physical processes that shaped them; every survey for exoplanets contains biases that limit the types of exoplanets detected. Thus, observed populations must be corrected for incompleteness and biases to obtain intrinsic occurrence rates, which can subsequently be directly compared to models. The importance of occurrence rates is illustrated by the case of hot Jupiters, which are relatively easy to detect in transit because of their large radii and short periods, but which are intrinsically rare \citep[e.g.,][]{Howard2012, Zhou2019, Dattilo2023} -- a conclusion that dramatically alters our understanding of planet formation. 

The data used to infer occurrence rates for exoplanets must be well-characterized, with careful sample selection, uniform observations and detections, and quantified biases. NASA's Kepler prime mission, which flew from 2009-2013 \citep{Borucki2010}, exemplified this observation strategy and thus remains the gold standard sample for exoplanet occurrence rates, even more than 10 years after the conclusion of the prime mission. 

The primary mission goal of Kepler was to measure the occurrence rate of Earth-like planets around Sun-like stars using the transit method. Because of the failure of a second reaction wheel at the end of the prime mission and unexpected levels of stellar astrophysical noise, Kepler was not sensitive to Earth analogs \citep{Lissauer2023}.  However, the Kepler observations facilitated a swath of occurrence rate calculations, both for exoplanets closer to the Kepler sensitivity range as well as extrapolations out to the Earth-analog regime \citep[e.g.,][]{Petigura2013, Foreman-Mackey2014, Burke2015, Kunimoto2020, Bryson2021, Bergsten2022}.

Occurrence rate calculations over the last 15 years, with Kepler or with other techniques, have been varied in their samples, observation strategies, and methodologies, but they all have one feature in common: they have almost all focused on exoplanets in single star systems, either explicitly or implicitly (with at least one exception: \citealt{Hirsch2021}). However, binary stars (and higher-order multiples) are a significant fraction of the field star population. Half of Sun-like FGK stars have a stellar companion \citep{Raghavan2010, Offner2023}, with multiplicity fractions increasing to higher masses and decreasing to lower masses. Even the lowest-mass stars have multiplicity fractions of $\sim$ 20-25\% \citep[e.g.,][]{Winters2019}. 

Binary stars complicate stellar characterization, planet detection, and survey characterization, and so demographic studies, including occurrence rates, typically attempt to exclude binaries from their stellar samples. This is a difficult task, because directly and conclusively ruling out multiplicity for a sample of tens or hundreds of thousands of stars is generally observationally infeasible. This was true during the construction of the Kepler Target Catalog (KTC; \citealt{Batalha2010}), and is still true now even with the commonly-used and reliable binary star metrics available in Gaia Data Releases 2 and 3 (e.g., the Renormalized Unit Weight Error, or RUWE; \citealt{Gaia2016, Gaia2018, Lindegren2021, Gaia2023}). Some previously known wide-separation binaries were de-prioritized or excluded from the Kepler target list because of a priori knowledge of flux dilution that precluded small planet detection \citep{Batalha2010}, but small-angular-separation binaries (angular separation $\rho \ll 4''$, which is the size of a Kepler pixel) were still observed, and some had planet detections.

As Kepler Objects of Interest (potential planet candidates; KOIs) were published, the scope of binary contamination in the KOIs began to be quantified by extensive follow-up of the KOIs with high-resolution imaging via speckle interferometry and adaptive optics (AO) imaging \citep[e.g.,][]{Howell2011, Lillo-Box2012, Lillo-Box2014, Wang2014, Wang2015, Kraus2016, Furlan2017, Ziegler2017, Ziegler2018}. The planets and planet candidates that were ultimately discovered to be in binary star systems produced the first large sample of planets in binary stars. Studies of this sample have revealed hints that the population of planets in binary star systems may differ substantially from those in single stars. 

For example, planets in binaries appear to be less common than those in single stars, a feature first observed in the Kepler sample and since replicated in many other samples \citep{Wang2014, Kraus2016, Ziegler2018, Ziegler2020, Lester2021, Clark2024}. More recently, \citet{Sullivan2023} and \citet{Sullivan2024b} explored the properties of planets in binary stars and found that the presence of the radius valley and the relative number of super-Earths and sub-Neptunes appears to be binary-separation-dependent, with sub-Neptunes suppressed in small-separation binaries. 

However, all of these results were analyzed using raw planet counts, rather than drawing conclusions from an underlying inferred occurrence rate that has been calculated using bias-corrected survey statistics. The process of discovering binary stars within the $\sim$8000 KOIs is labor-intensive, but doable with enough effort and telescope time, meaning that it is possible to assess basic demographics for planets in binary star systems using planet counts without corrections for survey incompleteness, but it is not possible to directly compare to single star populations or to theory.

In contrast, occurrence rates present an unbiased reflection of the true planet statistics, but to correct the intrinsic survey biases a well-characterized stellar sample is required. For single stars this is already a substantial effort, but binary stars are even more complicated because it is necessary to measure not only the individual stellar properties (which can be biased because of contaminating light from the other stellar component; \citealt{Furlan2020, Sullivan2022b}), but also the binary properties like separation and mass ratio. Uniform stellar characterization was undertaken several times for the Kepler sample, starting with the Kepler Input Catalog (KIC; \citealt{Brown2011}) and continuing to efforts incorporating Gaia data \citep{Berger2018, Berger2020stars}, but these efforts did not include additional observations to detect binaries in the KTC. Detailed high-resolution imaging surveys like the efforts undertaken for the KOIs are not possible for the full sample of 200,000 stars in the KTC, meaning that an alternative mode for identifying and characterizing binary stars must be developed before it is possible to accurately calculate occurrence rates for planets in binary star systems.

In this work, we present the first occurrence rates for transiting planets in binary star systems. To create a parent population of binary stars, we assembled a sample of high-likelihood binaries observed by Kepler and identified from Gaia DR3 data. We then characterized the sample of high-likelihood binaries using a Monte Carlo statistical approach to produce posterior distributions for the stellar and system properties (temperatures, radii, luminosities, masses, binary separation). To carry through statistical uncertainties to the final occurrence rate calculation, we generated 500 stellar catalogs using a mixture model incorporating uncertainty in binary status, uncertainty in planet host star, and uncertainty in the stellar parameters. To assemble the candidate binaries we leveraged the fact that most binary stars with elevated RUWE in Gaia DR3 have angular separations $\lesssim$0.2'' \citep{Dodd2024, Sullivan2025a}, meaning that they were fully unresolved in Kepler observations. Thus, we were able to generate a sample of candidate binary systems that is relatively unbiased to exclusion from Kepler observations. This is in contrast to wider binaries, which would have been spatially resolved and potentially excluded from the survey due to egregious flux dilution.

We generated 500 realizations of the binary catalog and evaluated the survey completeness for each, taking into consideration the impact of flux contamination on planet detectability for  individual systems, then used the completeness to calculate the occurrence rates for each catalog. We combined all 500 occurrence rate estimates to obtain one final probabilistic occurrence rate calculation for planets in small-separation binary star systems. Section \ref{sec:sample} describes the sample selection for stars and planets, including stellar catalog generation; Section \ref{sec:occ} describes the components of the occurrence calculation, including the modifications we made to each calculation to compensate for stellar multiplicity; Section \ref{sec:rates} presents the occurrence rates; and Section \ref{sec:discussion} interprets our results and summarizes the paper.

\section{Sample Selection} \label{sec:sample}
\subsection{Stellar Sample}

\begin{figure}
    \includegraphics[width=\linewidth]{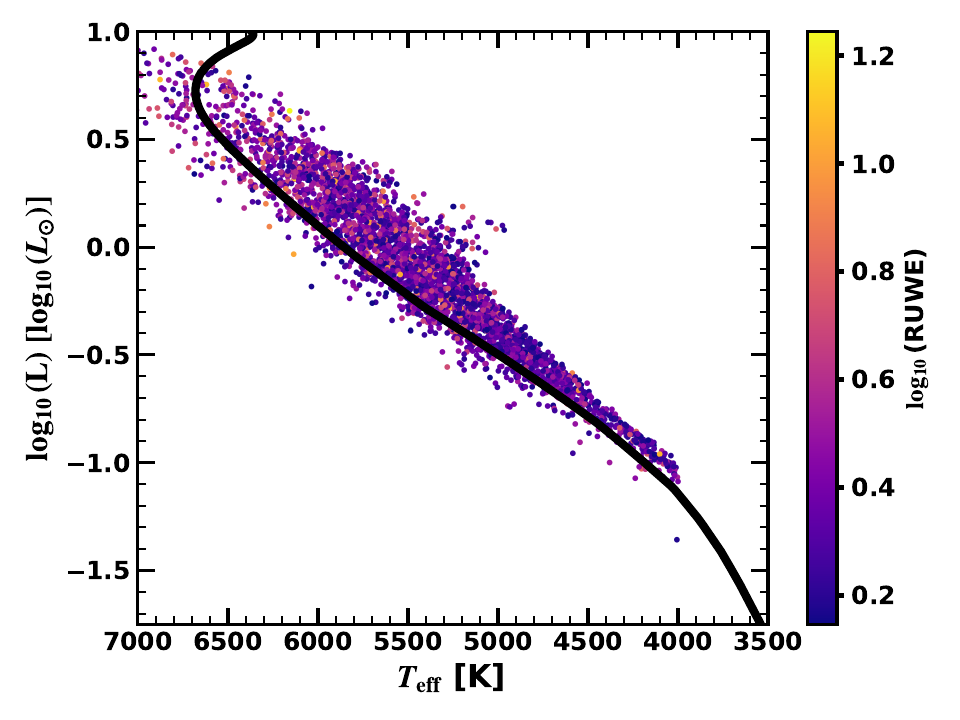}
    \caption{An HR diagram of the full stellar sample of candidate binaries, with an evolutionary model from MIST overplotted (black line) and color-coded by $\log_{10}$(RUWE) from Gaia. The candidate binaries almost all fall above the main sequence defined by the evolutionary model, as expected for a sample identified using astrometric excess noise and HR diagram offsets.}
    \label{fig:hrd}
\end{figure}

To construct a stellar catalog for the occurrence rate calculation, we needed to identify candidate binary systems, determine distributions for the stellar properties, then generate realizations of the catalog to carry through the probabilistic occurrence calculation. We selected an initial candidate binary system sample (the ``parent'' sample) using the catalog of \citet{Berger2018} and \citet{Berger2020stars}, who included flags for stellar multiplicity based on several binary identification processes, including Gaia metrics, HR diagram offset, and high-contrast imaging. We selected systems only with uniform identification techniques: main sequence offset and Gaia metrics. We supplemented the \citet{Berger2018} and \citet{Berger2020stars} binary flags by cross-matching the candidate binaries with Gaia DR3 \citep{Gaia2023}, where we used the Renormalized Unit Weight Error (RUWE) metric (a measure of the noise in the Gaia astrometric solution) to provide another constraint for selecting high-likelihood binaries by selecting only sources with a \texttt{RUWE $>$ 1.4}, choosing the cutoff following the suggestion of \citet{Lindegren2021}. 

We selected systems only with elevated Gaia RUWE, some of which also had main sequence offsets, both of which are uniform follow-up methods. Main sequence offset is sensitive mostly to smaller separation (photometrically unresolved), equal-mass binaries, while elevated RUWE is typically caused by photocenter motion of small-angular-separation binaries, and is more agnostic to mass ratio. We ensured that our entire sample had elevated RUWE, so we expected the stars to mostly be small-separation and with a range of mass ratios.

We also followed the same process as \citet{Dattilo2023} to further ensure a high-quality stellar sample, removing flagged evolved stars, non-FGK stars (T$_{\rm eff} < 4000$ K or T$_{\rm eff} > 7300$ K, and stars with poor Gaia astrometric fits (parallax error $>$ 10\%). The sample is shown in an HR diagram in Figure \ref{fig:hrd}, including an evolutionary track at 2 Gyr from the MESA Isochrones and Stellar Tracks (MIST) models \citep{Paxton2011, Paxton2013, Paxton2015, Choi2016, Dotter2016}.

\begin{figure*}
    \includegraphics[width=0.5\linewidth]{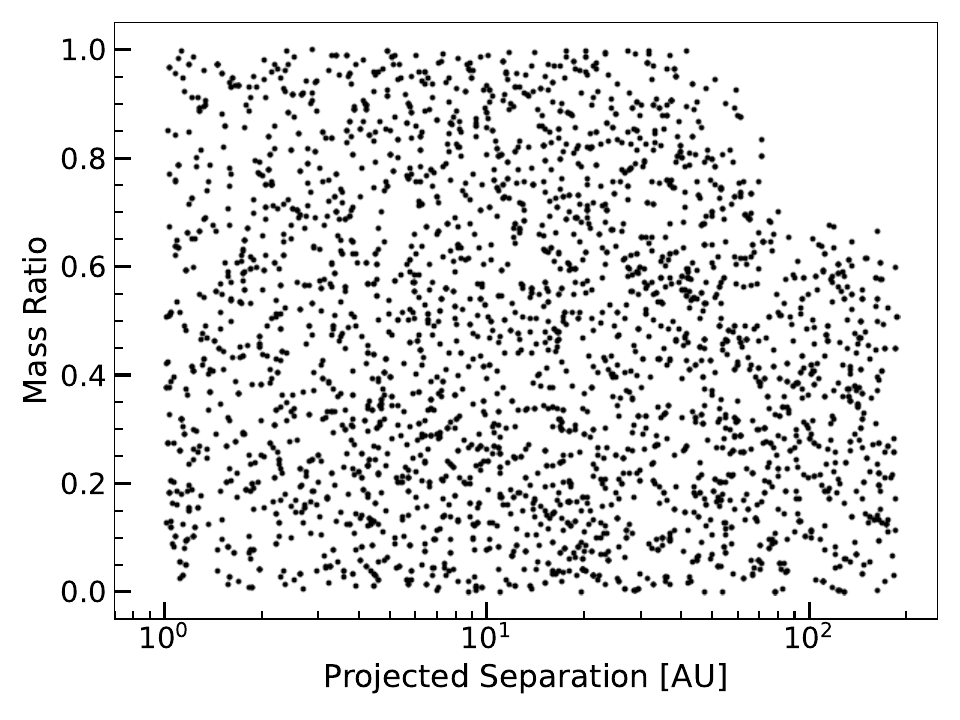}
    \includegraphics[width=0.5\linewidth]{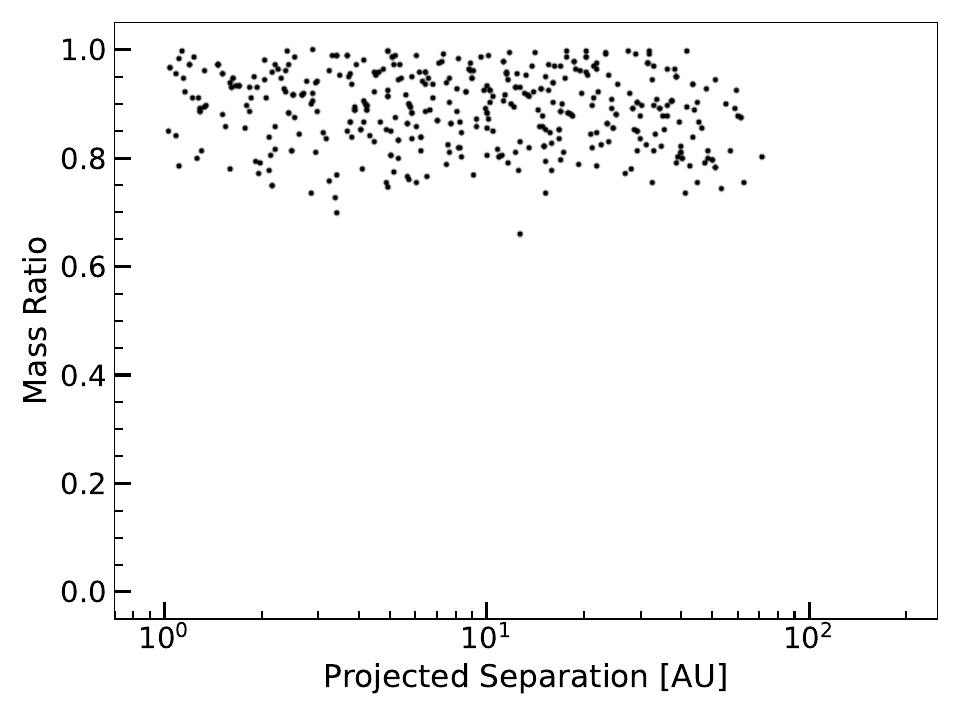}
    \caption{The mass ratio versus projected separation for a sample of binaries drawn from a MOLUSC run for Gaia DR2 2086403376201149184, a randomly selected candidate binary with RUWE = 1.61. We drew randomly from the MOLUSC samples to calculate our simulated binary properties for each target and create a catalog. Left: The unfiltered distribution is mostly uniform, because MOLUSC produces draws with a wide range of RUWE values. Right: The distribution filtered to only include systems with estimated RUWE values within 25\% of the true RUWE value. The RUWE filtering plus MOLUSC provides a strong constraint on mass ratio and projected separation.}
    \label{fig:molsuc}
\end{figure*}

To determine the binary star properties of separation and mass ratio, we used the Multi Observational Limits on Unseen Stellar Companions (MOLUSC; \citealt{Wood2021}) package to infer the distribution of possible binary separations and mass ratios using the RUWE of each target. MOLUSC can take in a variety of information about a system, including results from radial velocity monitoring, high-resolution imaging follow-up, and Gaia metrics like the system's RUWE value, and MOLUSC can be used with any combination of observed data sets as long as the target star has a measured mass. After inputting the relevant data, MOLUSC runs a user-determined number of injection-recovery simulations over the parameter space of possible binary properties to assess which binary systems would have been detected given the observed constraints and which would have gone undetected, then reports the parameter space in which a binary companion would not have been detected by the observations. In the case of the candidate binary systems, we only input the system RUWE and the primary star mass as determined from \citet{Berger2020stars}, assuming that the secondary star introduced a minimal bias to the measured primary star mass. We ran 10,000 injection-recovery tests over the full parameter space, and took any surviving MOLUSC results (i.e., binary configurations that would have been undetected by Gaia given the star's RUWE) to be a possible population from which to draw our statistical sample. From the surviving MOLUSC samples for each target (i.e., systems that could have produced the observed signal without detection from Gaia), we downselected to only include systems with predicted angular separations $<$ 0.2'', based on the results of \citet{Sullivan2025a}, who found that binary systems with elevated RUWE in Gaia typically had angular separations $\lesssim$0.2''. We also restricted the MOLUSC results to samples with predicted RUWE values within 25\% of the measured value. 

Figure \ref{fig:molsuc} shows the projected separation-mass ratio distribution for the selected MOLUSC samples for a random target. As apparent in the figure, the RUWE alone gives a degenerate solution to the mass ratio-projected separation relation, with a section of parameter space ruled out because those binaries would have been identified by Gaia if they existed. From the filtered sample of MOLUSC results, we randomly selected 3000 samples to proceed with our analysis for computational efficiency. For each star we used the reported mass ratio (q) and projected physical separation ($\rho$) distributions from the MOLUSC samples, alongside the measured mass M$_{\rm pri}$ from \citet{Berger2020stars} to infer stellar properties for the secondary star, such as mass, temperature, and radius. We also calculated the planet radius correction factor (PRCF; \citealt{Ciardi2015, Furlan2017}) for planets given a primary or secondary star host. The PRCF is a multiplicative factor that corrects an observed planet radius for the effects of flux dilution.

To calculate the occurrence rates we used multiple realizations of the analysis for each star. Thus, to generate the complete stellar sample we generated 500 catalogs, drawing from the distributions in mass ratio and binary separation, as well as the corresponding stellar parameters, such as effective temperature ($T_{\rm eff}$) and radius. For each catalog we assumed 10\% single star contamination (which were excluded from subsequent analysis) based on comparisons to the California Kepler Survey sample \citep{Petigura2022}, and 80\%/20\% primary versus secondary star planet hosts, which determined which set of stellar parameters we used.  We do not know if planets form more frequently around FGK primary stars versus their companions.  However, there are examples in the discovery catalogs of planets orbiting secondary stars (e.g., \citealt{Gaidos2016}, Burns-Watson et al. in review). Assuming that Kepler candidates always orbit the primary would introduce biases in the derived occurrence rates.  To mitigate this issue, we assume an 80\%/20\% ratio of planets orbiting the primary vs. secondary, loosely informed by \citet{Gaidos2016} who found that $\sim$90\% of planets om M dwarf binaries orbit the primary.  We were more conservative due to the higher stellar masses of the parent sample. We also tested the calculation using a 50/50 mixture of primary versus secondary star hosts for the planets and found that our results did not change. The cleaned catalog of parent binaries, before making additional quality cuts from the Kepler data, had 5157 systems.

For each stellar catalog, we also made a series of cuts based on the quality of the Kepler data associated with each target. We removed targets flagged by the Kepler Noisy Target List; all stars with $R_{\star} > 1.25 R_{\odot}$; stars with undefined limb darkening coefficients; stars with undefined duty cycles, a drop in duty cycle of $>$ 30\% from removal of transits or a duty cycle $<$0.6; stars with data span $<$ 1000 d; and stars with a flag indicating that the transit fit timed out. On average, the final catalogs typically had $\sim 2000-3000$ stars, with varying numbers depending on how many targets were flagged as random single stars and the data quality for the stars in the remaining sample. Table \ref{tab:stars} shows the full stellar catalog with the derived and measured values for each system, including the widths of the distribution from which we drew for each realization of a test catalog.

\subsection{Planet Sample}
We assembled a planet sample by downloading the full table of Kepler Objects of Interest (KOIs) from Kepler data release 25 \citep[DR25;][]{Thompson2018} from the NASA Exoplanet Archive\footnote{\url{https://exoplanetarchive.ipac.caltech.edu/cgi-bin/TblView/nph-tblView?app=ExoTbls&config=cumulative}; accessed 7 April 2025.}. For the occurrence calculation, we removed any planets flagged as False Positive (FP), and retained planets flagged as Planet Candidate (PC) or Confirmed Planet (CP). For each of the 500 catalog iterations we restricted the sample to planets belonging to stars in the relevant stellar catalog. The planet catalogs typically contained $\sim$30 PCs/CPs and $\sim$20 FPs. 

\begin{figure}
    \includegraphics[width=\linewidth]{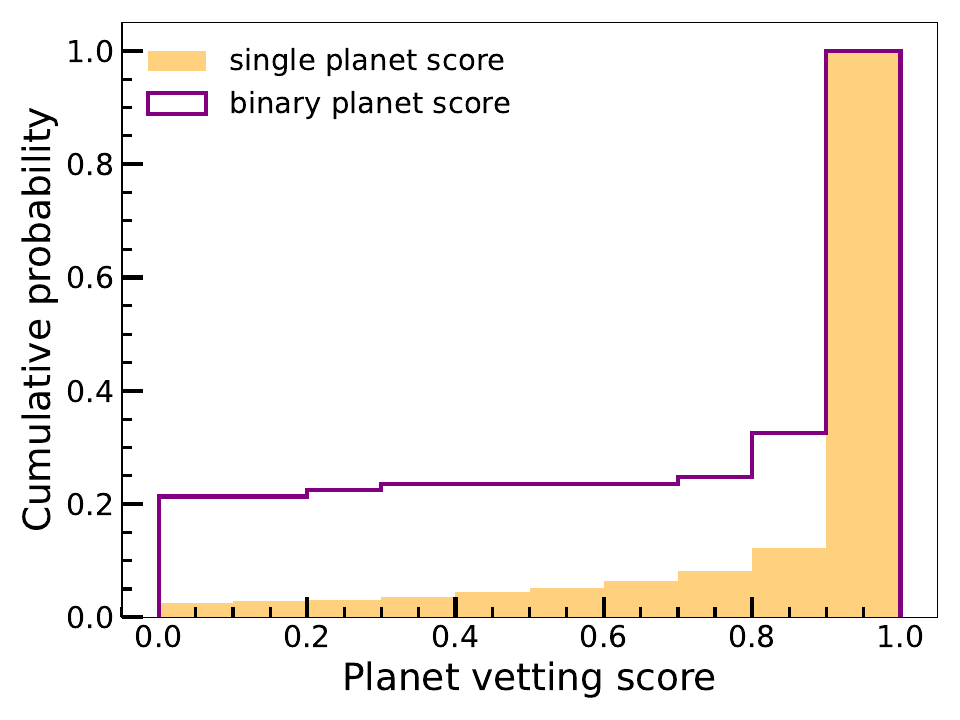}
    \caption{Cumulative distribution functions of the vetting score for planets in binaries (purple) versus single stars (orange). The planets in binaries tend to have somewhat lower vetting scores than the planets in single stars, but the difference is not statistically significant ($< 2\sigma$).}
    \label{fig:vetting_score}
\end{figure}

To investigate whether the planet sample and confidence for the binaries was similar to those of single stars, we compared the vetting scores for the planet sample in binaries versus the single stars. Figure \ref{fig:vetting_score} shows a cumulative distribution function of the vetting scores for the binaries and single stars. The planets in binaries systematically have somewhat lower vetting scores than the planets in single stars. However, a Kolmogorov-Smirnov statistical test shows a $< 2\sigma$ difference between the two distributions, indicating that there is no statistically significant difference between the vetting scores of planets in binaries versus single stars.

\section{Components of Occurrence Rate Calculations} \label{sec:occ}
To adequately measure an exoplanet occurrence rate, the Kepler DR25 catalog of viable transiting exoplanet candidates must be corrected for biases. For example, some transit events may have been missed by the detection pipeline (lack of completeness in some regimes), while others may have been incorrectly dispositioned as FP or PC. We attempted to compensate for these survey biases using corrections for both the completeness and the reliability of our sample. For both sets of corrections we followed the methodology of \citet{Dattilo2023} and \citet{Bryson2020}, with modifications for binaries as described below.

\subsection{Completeness Model}
\subsubsection{Pipeline Completeness}

\begin{figure}
    \includegraphics[width=\linewidth]{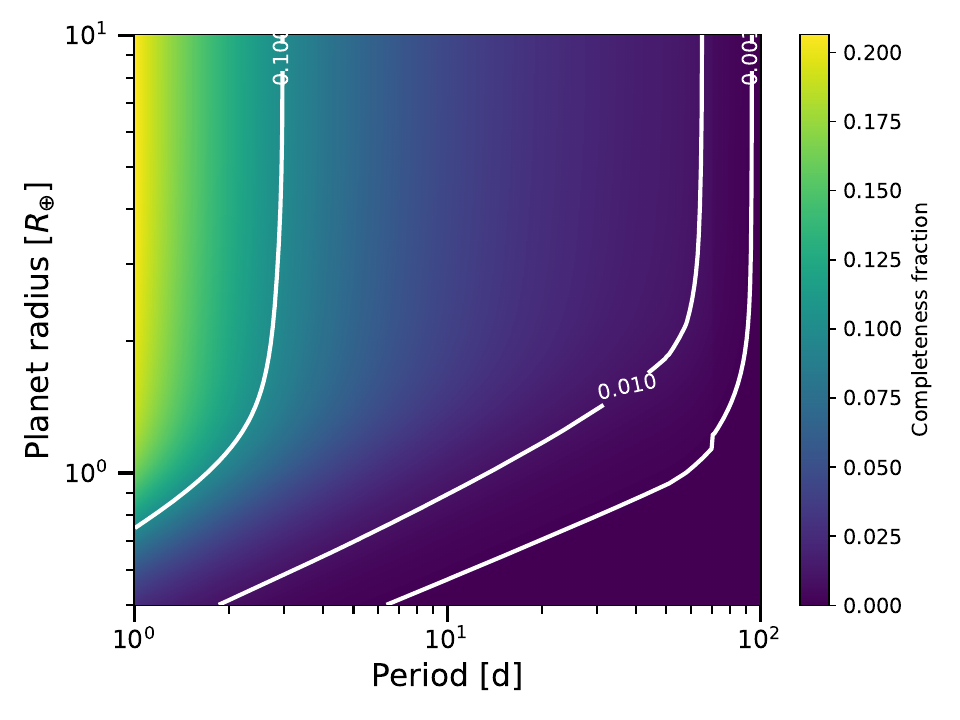}
    \caption{Mean completeness map for the binary stellar population, incorporating geometric transit probability. The contours correspond to completeness of 10\%, 1\%, and 0.1\%. The completeness on average is lower than for single stars, and is shifted upward in planet radius relative to single stars.}
    \label{fig:completeness}
\end{figure}

The detection completeness quantifies the fraction of true transiting planets that were detected at or above a given signal threshold (chosen to be 7.1$\sigma$ in order to yield less than one statistical false alarm over the full survey) by the Kepler pipeline. Detection completeness characterizes the regime of observations where there is high completeness (most/all of real planets are detected) or low completeness (most/all real planets are missed or unobservable). The detection completeness is a function of both period and transit signal-to-noise ratio, which is reported as the Kepler Multiple Event Statistic (MES). 

We calculated detection completeness for each stellar catalog using a modified version of the KeplerPORTS\footnote{\url{https://github.com/nasa/KeplerPORTs}} code, which was first presented in \citet{Burke2017} and subsequently modified by \citet{Bryson2020}. We used the updated version from \citet{Bryson2020}, with modifications for binaries as described below. 

Detection completeness is calculated for each star in the sample individually, then summed to produce a final detection completeness map for a given sample. The geometric transit probability is also calculated at this stage and included in the final detection completeness. For our methodology we calculated detection completeness independently for each of our 500 randomly generated stellar catalogs. The mean completeness map is shown in Figure \ref{fig:completeness}.

Because the detection completeness involves the sensitivity to transiting planets at a given MES and period, we needed to compensate for the presence of binaries at any place in the calculation where there was a dependence on stellar mass or radius. The assumed stellar mass and radius change depending on whether the planet is around the primary or secondary star, and also alter the measured MES. We made modifications to KeplerPORTS for the geometric transit probability, the transit duration, and the transit depth.

The geometric transit probability is the probability that a system will be in a configuration that will produce a transit. Thus, it is dependent on the stellar radius. In a binary system, the geometric transit probability is dependent on the radius of the host star of the planet, so we modified KeplerPORTS to consider only the radius of the planet host star when calculating the geometric transit probability. The geometric transit probability also depends on the planet's semi-major axis, which is calculated using Kepler's third law using the mass and period. Thus, we also substituted the true host star mass rather than the apparent mass into the geometric transit probability.

The transit duration is the length of time it takes a planet to cross the projected surface of the stellar disk. It is also dependent on the stellar mass and radius, which we ensured were the mass and radius of the assigned host star.

Multiple bias corrections are implemented in KeplerPORTS, parameterized as a function of the MES. Completeness contours are calculated as a function of P and $R_{p}$. Every grid point in the completeness contour must be mapped to a specific MES value. This is done via the 1-$\sigma$ depth function (itself a function of P; \citealt{Burke2017onesigma}). The 1-$\sigma$ depth function is derived from light curve properties, and is defined as the transit depth for a star that results in MES = 1. Transit depth is defined as $\delta \equiv \left(\frac{R_{p}}{R_{\star}}\right)^{2}_{\rm true}$ but because our targets are binaries with flux dilution, to achieve this we needed to convert the transit depth considered by the 1-$\sigma$ depth function from $\left(\frac{R_{p}}{R_{\star}}\right)^{2}_{\rm obs}$. In a binary system the transit depth is diluted \citep{Ciardi2015}, meaning that the sample is less sensitive to small planets than if it was composed of single stars. To address the flux dilution, we modified all 1-$\sigma$ depth functions by multiplying by the ``planet radius correction factor'' or PRCF. The PRCF is a simple mathematical factor that describes the flux dilution in the Kepler bandpass and can be calculated using the contrast between the two binary components, as well as their radii:

\begin{equation}
    f_{\rm pri} = \frac{R_{\star, {\rm pri}}}{R_{\star, {\rm total}}} \sqrt{1 + 10^{-0.4 \Delta m_{Kep}}}
\end{equation}
\begin{equation}
    f_{\rm sec} = \frac{R_{\star, {\rm sec}}}{R_{\star, {\rm total}}} \sqrt{1 + 10^{0.4\Delta m_{Kep}}}. \\
\end{equation}

The remainder of the calculations, including the baseline detection efficiency model (stellar radius, CDPP slope, and orbital period dependencies and their tabulated corrections, as discussed below), are dependent on the apparent (or observed) radius of the star, not the true radius of the star. For those calculations, we used the published (observed) \citet{Berger2018} stellar radius instead of the corrected host star radius.

\begin{figure}
    \includegraphics[width=\linewidth]{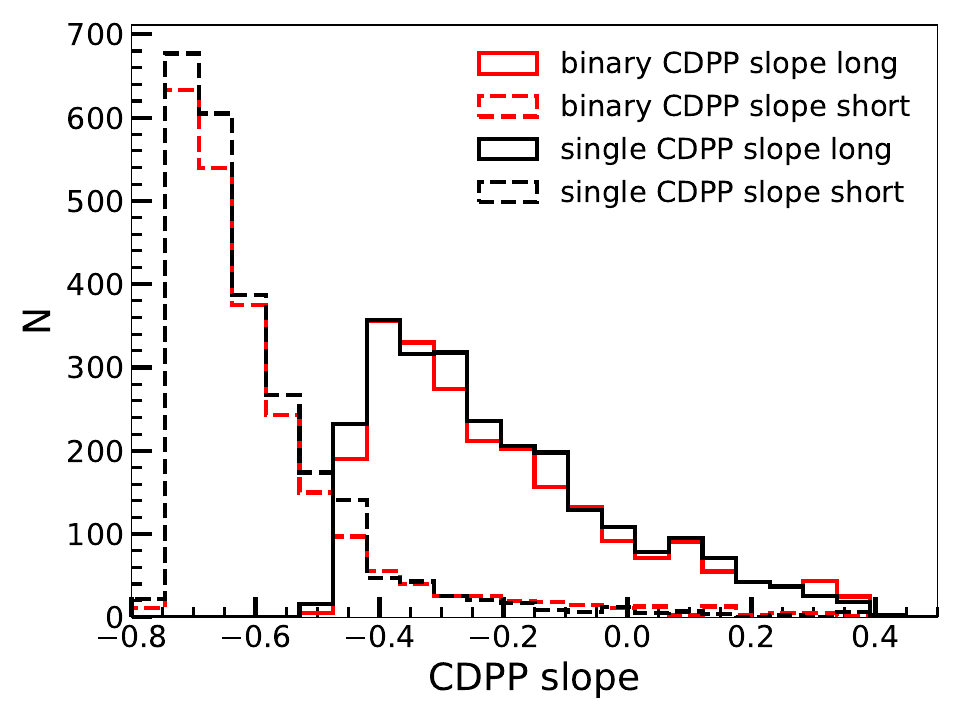}
    \caption{CDPP slope compared for both short (60s, dashed line) and long (30 m, solid line) cadence data for binaries (red) and single stars (black). There is no significant difference between the noise properties of the binary stars versus the single stars in either case, per a Kolmogorov-Smirnov test.}
    \label{fig:cdpp_slope}
\end{figure}

To test the assumption that the binary and single stars had the same light curve properties (and thus that the KeplerPORTS lookup tables for relationships between stellar properties and noise properties were still valid  for the candidate binaries), we compared the combined differential photometric precision (CDPP) slope values between the two samples, and investigated both the short (60\,s) and long (30\,m) cadence data. Figure \ref{fig:cdpp_slope} shows histograms of the CDPP slopes for the four cases we considered, and demonstrate that the distributions are identical in both the short and long cadence data. We validated that observation by performing a Kolmogorov-Smirnov test on the long and short cadence samples and found $< 1\sigma$ difference between the binaries and singles in both cases, indicating that their noise properties are identical.

\subsubsection{Vetting Completeness}
The vetting completeness is used to assess the likelihood of a threshold crossing event (TCE) being correctly dispositioned as a PC rather than a FP. We used the same method to calculate vetting completeness as described in \citet{Bryson2020}, and describe it briefly here. 

Vetting completeness is calculated by analyzing the number of PCs detected from injection-recovery tests performed  in period-expected MES space \citep{Christiansen2017, Burke2017}. The results from the injection-recovery tests are binned, and we treated the fraction of recovered PCs in each bin as a binomial rate. We then fit the surface of the grid using an MCMC analysis with \texttt{emcee} \citep{Foreman-Mackey2013}, assuming a simple logistic function in period and a rotated logistic function in period-MES space. 

We assumed that the presence of a small-separation binary star did not significantly impact the vetting completeness results. This may not be the case, given that the vetting completeness depends on the results of injection-recovery tests that were performed assuming that each target was a single star. However,  the injection-recovery tests are performed in period-MES space, which only depends on the properties of the lightcurve and therefore is independent of stellar properties. Thus, we do not expect any changes to the injection-recovery results caused by the stellar multiplicity, other than the appropriate scaling for MES.

In theory, the presence of a companion star may change the results of injection-recovery tests that were originally used to calibrate the detection completeness calculation \citep{Christiansen2017}. However, it was not computationally feasible to run our own pixel-level injection-recovery tests. Additionally, because none of our sources were resolved binaries in the Kepler observations, a correction to the calculation via alterations to the 1-$\sigma$ depth function was sufficient to alter the sensitivity of the detection completeness for the binaries. Similarly, we assumed that all other pixel-level or pipeline-level modeling and calculations were consistent regardless of the binary star nature of the host. Future work will investigate this assumption in more detail.

Each factor in the completeness calculation (geometric transit probability, detection completeness, and vetting completeness) was treated as independent and multiplied together to give the completeness for each star, then summed to give a total completeness over the full stellar sample. 

\subsection{Reliability Model}
\subsubsection{Astrophysical False Positives}
Astrophysical false positives encompass any sources that might produce a light curve feature that resembles an exoplanet transit. The dominant source of astrophysical false positives is binary systems, often identified because of their v-shaped transits or an offset in the Kepler pixel because of a background source. However, we assumed all our targets had angular separations much less than the size of a Kepler pixel, and so would not have been identified via an offset during the astrophysical false positive identification stage. If the targets did have larger separations, they would have been resolved by Gaia and not included in our candidate binary star sample, as well as being flagged as an FP by Kepler because of a potential offset. We assumed 100\% accuracy for all targets and so did not correct for astrophysical false positives. The main focus of the astrophysical false positive step is to identify stars in binary systems, so many of our targets may be flagged as a spurious astrophysical false positive. Thus, we assumed that the astrophysical false positive assessment was not valid for our sample, and did not include it in our calculations.

\subsubsection{False Alarm Reliability}
The false alarm reliability assesses how likely a false alarm (FA) is to be contaminating the planet sample. FAs are produced by instrumental artifacts or stellar noise in a lightcurve, and represent a different class of detection than an astrophysical FP. To calculate the FA reliability we followed \citet{Bryson2020}, who describe the methodology in detail, and who characterize its effects on occurrence rates in \citet{Bryson2020b}. 

Following \citet{Bryson2020}, we used a probabilistic approach to calculating the FA reliability for a given target (i.e., to assess whether a target is likely a FA). We followed a similar approach as for the vetting completeness, but with different inputs. To determine sensitivity to FAs, we used scrambled and inverted data in two separate tests \citep{Coughlin2016}, which does not contain any transit signals. Modifying the data by inverting or scrambling it ensures that there are no true transits or astrophysical false positives in the data, so any ``detections'' are the result of false alarms. We treated the fraction of signals detected in the scrambled data (and, separately, the inverted data) as a binomial statistics problem in period-expected MES space, and used MCMC inference to fit parameters to that space. We fit the data as a rotated logistic function. 

We assumed that the FA reliability was not changed between the single and binary star cases. This is because instrumental artifacts would be present in the data to the same extent regardless of the number of stars in the host system. In other words, flux dilution should not impact the FA rate from instrumental artifacts, except to the extent that brighter stars may have higher S/N transits than fainter stars. However, the presence of a detected FA signal is not altered by a higher flux baseline as produced by the composite observations of two stars. 

The reliability was calculated per planet, and was incorporated into the model by modeling each planet detection as a 2-dimensional Gaussian distribution with an amplitude and volume equal to the reliability value, and a bandwidth of 0.1 in log(P) and log(R).

\section{Occurrence Rates for Kepler Planets in Binaries} \label{sec:rates}

\begin{figure*}
    \includegraphics[width = \textwidth]{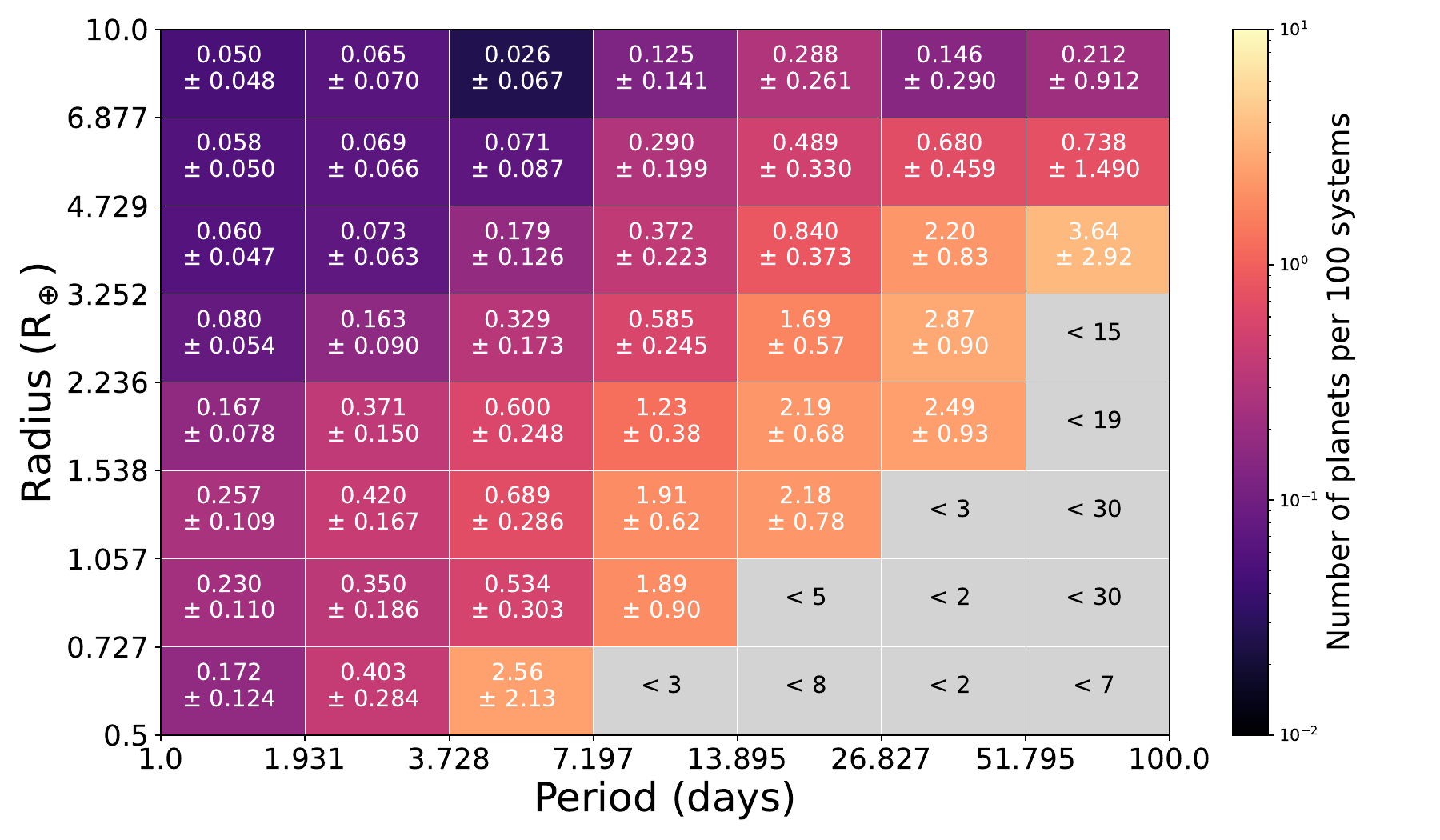}
    \caption{A map of occurrence in period-radius space for Kepler planets in binary systems with radii between 1-10 $R_{\oplus}$ and periods between 1-100 d. The gray boxes show regions of very low completeness where we assessed an upper limit. The colored boxes show the number of planets per 100 stars. The gray boxes show upper limits for regions with less than $0.1\%$ completeness.}
    \label{fig:occurrence_map}
\end{figure*}

We assembled a sample of high-likelihood small-separation binary systems, then calculated occurrence rates using completeness contours that were corrected for the impacts of stellar multiplicity. We performed 500 simulations of the occurrence calculation using a probabilistic framework incorporating uncertainties on the stellar properties, uncertainty in which star hosts the planet, and uncertainty in the stellar multiplicity (both the binary versus single nature of the source, and uncertainties in the binary properties of separation and mass ratio). We averaged the results of the simulations to produce a final occurrence map, and used the standard deviation of the simulations as the error on the occurrence.

Figure \ref{fig:occurrence_map} shows the occurrence in period-radius space, binned to a coarser grid than the original KDE estimate, for planets in binaries. The occurrence map spans from 1-100 d in period, and $1 - 10 R_{\oplus}$ in planet radius. The occurrence distribution broadly follows similar trends to single stars: low occurrence at large radii and short periods, high occurrence at long periods and small radii, albeit with large uncertainties because of the small sample. Two exceptions to the similarity are the lack of a steep drop-off in occurrence around 4 $R_{\oplus}$ (the radius cliff; \citealt{Dattilo2024}), and the overall lower occurrence rates relative to the single stars, as expected given previous observations of planets in binaries \citep[e.g.,][]{Kraus2016}. The gray boxes show upper limits for regions with less than $0.1\%$ completeness. 

To compare our integrated occurrence rates to single stars, we used the occurrence rates for Kepler single stars calculated by \citet{Dattilo2023}. The integrated occurrence rate over the full range of 1-100 d and 1-10 $R_{\oplus}$ (while excluding regions that are upper limits for either binaries or single stars) is 0.35$\pm$0.02 planets per system (NPPS) for planets in binaries, while it is 0.70$\pm$0.07 NPPS for planets in single stars \citep{Dattilo2023}, implying an occurrence rate of planets in close binaries that is $\sim 1/2$ that of single stars at 3.8$\sigma$ significance. In the more complete and better-sampled range of 1-50 d and 1-4 $R_{\oplus}$, the integrated occurrence rates are 0.226$\pm$0.005 NPPS for binaries and 0.534$\pm$0.022 NPPS for single stars \citep{Dattilo2023}, indicating $\sim$ 2.36 times more planets in single stars than in binary systems for this restricted period-radius range, or a nearly two-thirds reduction in the number of planets, with $\sim 11.4 \sigma$ significance.

\begin{figure*}
    \includegraphics[width=\linewidth]{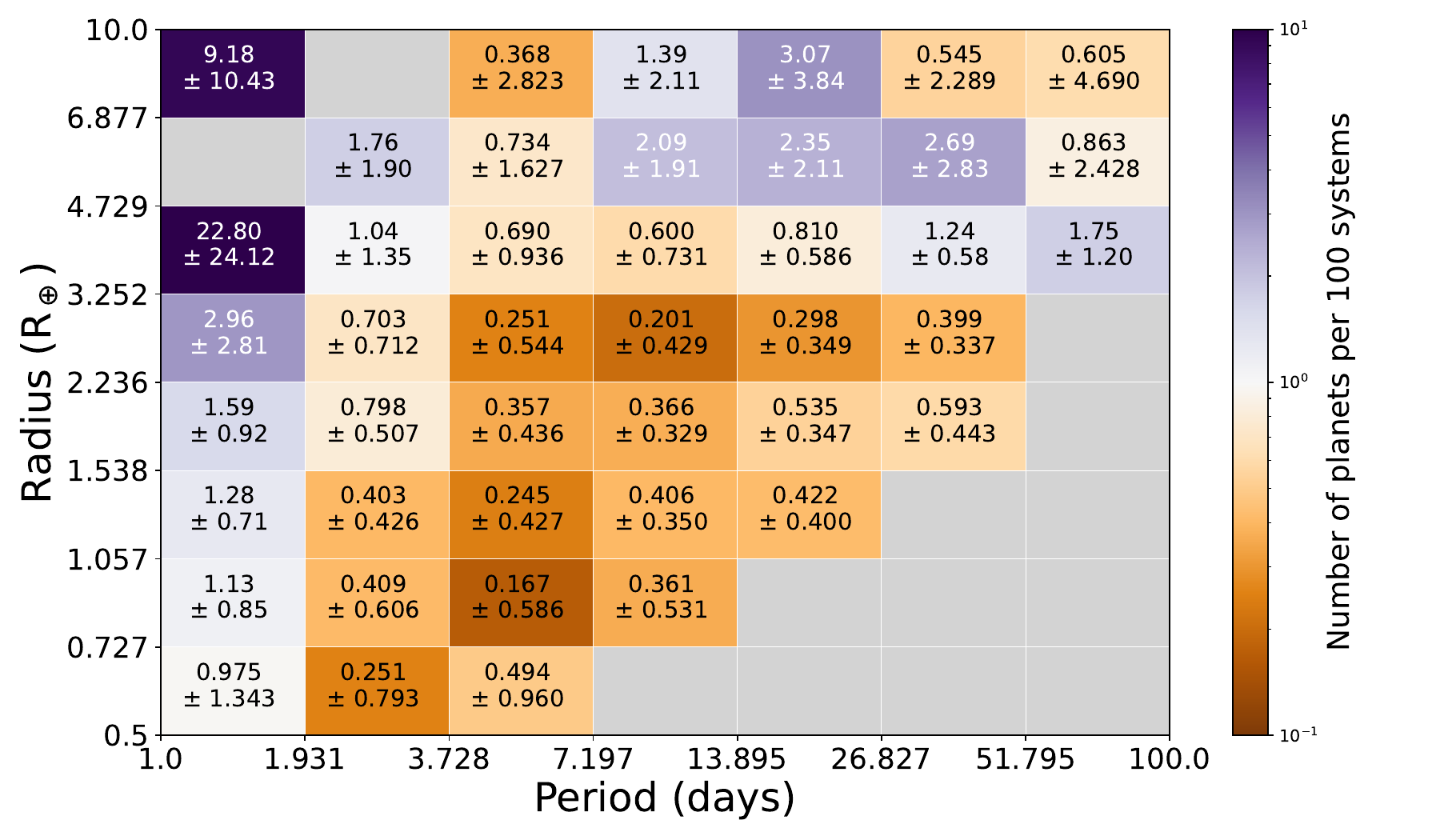}
    \caption{The ratio (binary occurrence/single occurrence) of occurrence rates for the same grid as Figure \ref{fig:occurrence_map}. A value less than one indicates that singles have higher occurrence, while a value greater than one indicates that binaries have higher occurrence. The gray boxes are regions where one of the two maps showed an upper limit. A substantial suppression of super-Earths and sub-Neptunes is apparent. The errors at large radii are on the order of the measurement, meaning that it is difficult to draw conclusions at radii larger than $\sim$ 4.7 $R_{\oplus}$. The sharp change in the ratio at 3.2 $R_{\oplus}$ is caused by the radius cliff in single stars, which reduces the planet occurrence until it is comparable to the planets in binaries. The binaries do not have an equivalent radius cliff.}
    \label{fig:ratio_map}
\end{figure*}

\begin{figure}
    \includegraphics[width=\linewidth]{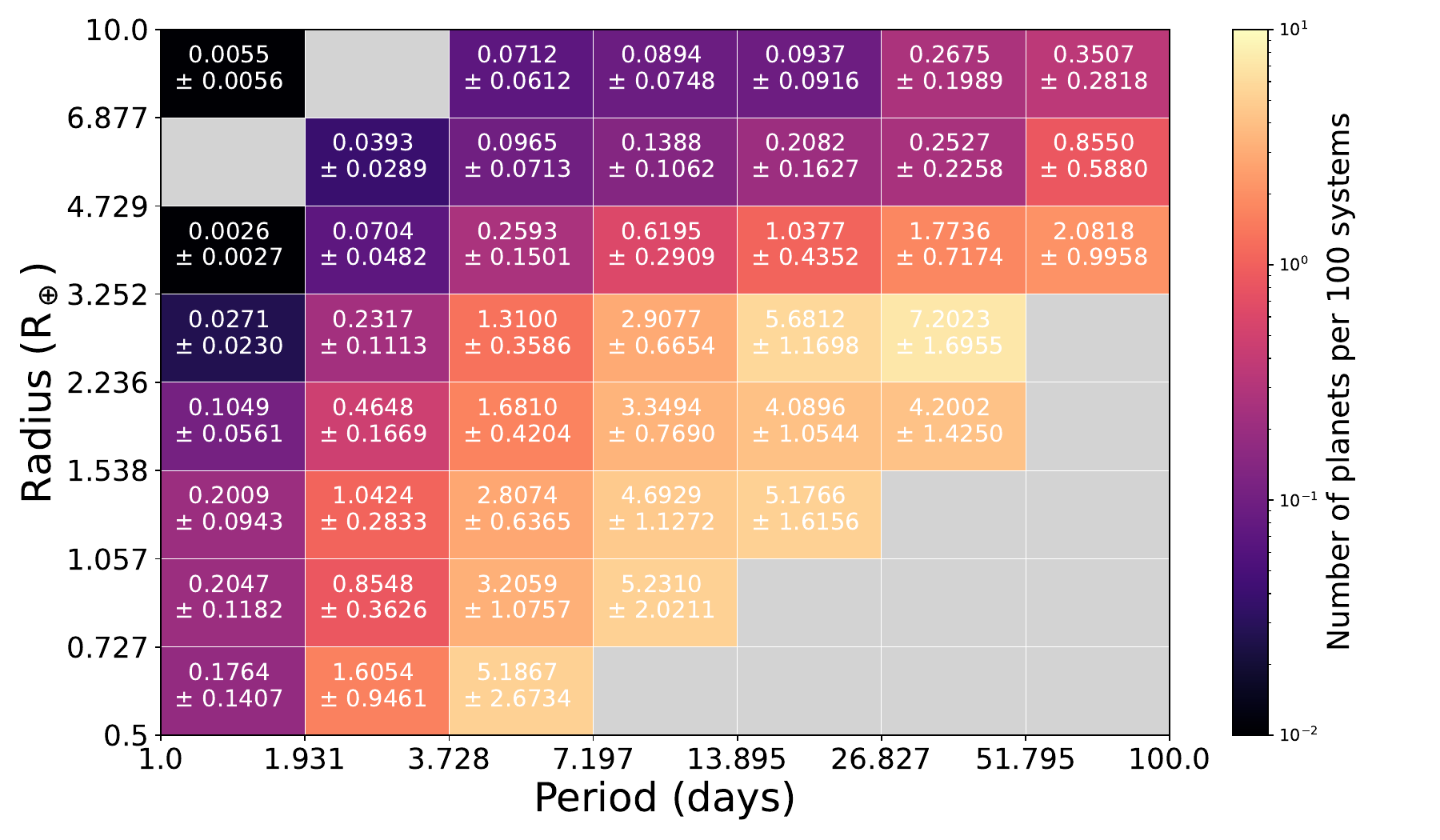}
    \caption{The occurrence of planets in single stars from 1-50\,d and 1-10\,$R_{\oplus}$, calculated using the methods from \citet{Dattilo2023}. The radius valley and radius cliff are both apparent.}
    \label{fig:single_map}
\end{figure}

To more easily compare between the single and binary stars, Figure \ref{fig:ratio_map} shows the grid of occurrence rates for planets in binaries divided by the grid of occurrence rates for planets in single stars, calculated from the occurrence maps in \citet{Dattilo2023} and shown in Figure \ref{fig:single_map}. Figure \ref{fig:ratio_map} shows several distinctive features. There is a substantial dearth of super-Earths and sub-Neptunes for the binaries compared to the singles, with most ratios being less than one. The sharp increase in the ratio around 3.2 $R_{\oplus}$ is because of the radius cliff in single stars, which reduces the occurrence of planets in single stars until it is comparable to the occurrence of planets in binaries at large radii. The very high ratios (and high occurrence rates for planets in binaries) at large radii and short periods are likely at least partially spurious because of small number statistics, and the errors are on the order of the measurement. The ratio map supports previous findings that small planets are suppressed in small-separation binary star systems relative to single star systems.

\section{Discussion}\label{sec:discussion}

\subsection{The Small Planet Radius Distribution}

One prominent feature of the exoplanet period-radius diagram is the radius valley, which falls around 1.7-1.8 $R_{\oplus}$ for single stars \citep[e.g.,][]{Fulton2017, VanEylen2018, Petigura2022, Ho2023}. In observations of binary stars, the radius valley appears absent in small-separation binary populations, which also show an apparent suppression of sub-Neptune exoplanets \citep{Sullivan2024b}. However, the \citet{Sullivan2024b} analysis was performed only for raw planet counts, so a direct comparison between the small planet radius distribution in single stars versus binary systems was not possible, since the observational biases differ between the single and binary star observations.

\begin{figure*}
    \includegraphics[width=\linewidth]{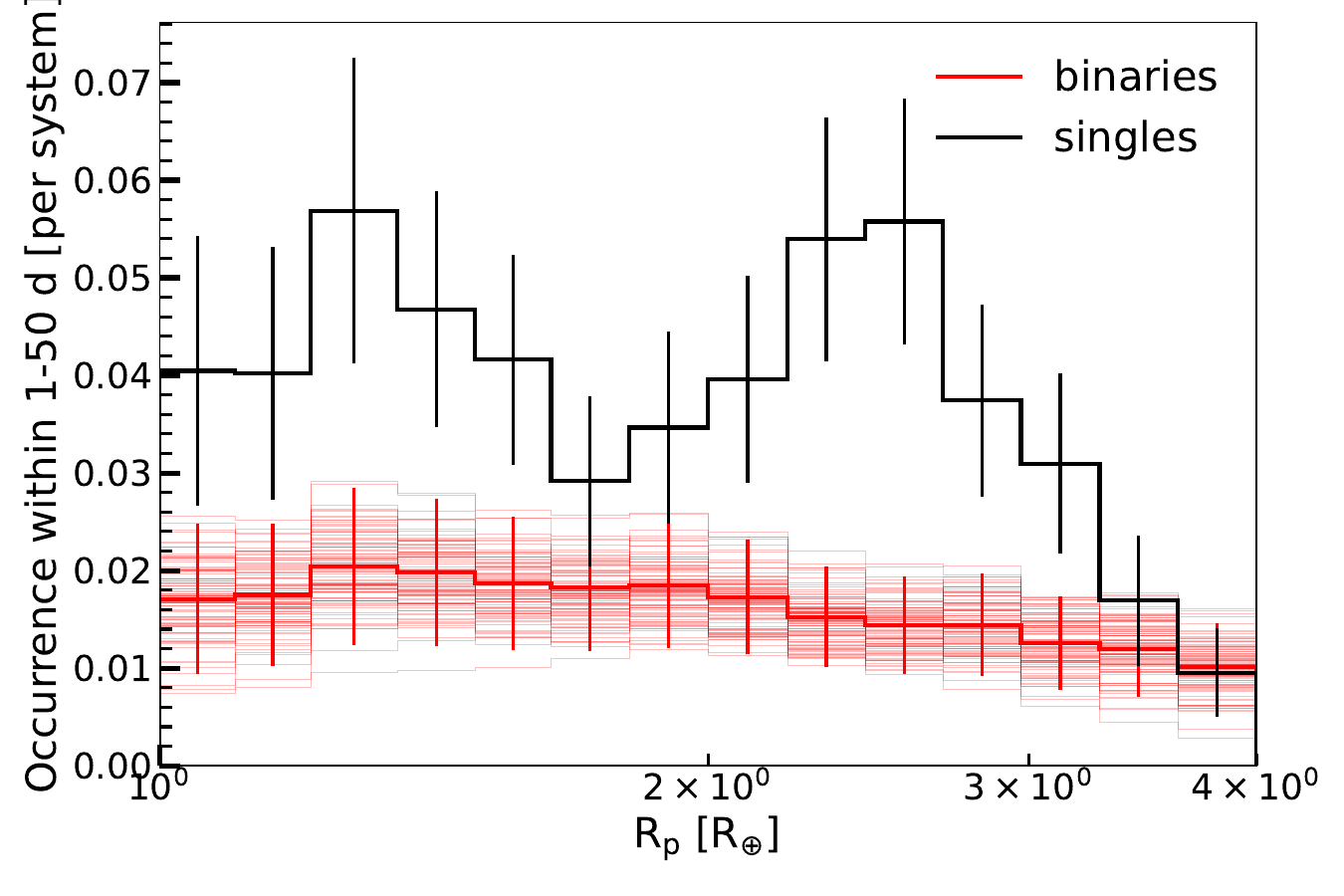}
    \caption{Radius distribution of small planets marginalized over periods fo 1-50 d for binary systems (red) and single stars (black; adapted from \citealt{Dattilo2023}). The thin red lines show scatter for the binary occurrence rate distribution as 100 distributions randomly drawn from the 500 original draws of the occurrence rate distribution, to show the general trends in the distribution. Although the total occurrence fluctuates for each draw, the shape of the distribution remains relatively consistent between draws. The occurrence for planets in binaries is substantially lower than the occurrence for planets in single stars, and does not show evidence for a radius valley or a radius cliff.}
    \label{fig:occ_rgap}
\end{figure*}

Figure \ref{fig:occ_rgap} shows histograms of the small planet radius distribution in occurrence space for both single stars and binary systems, marginalized over periods between 1-50 d: a period range with high completeness for both singles and binaries. The single star distribution shows a clear radius valley, relatively high occurrence rates, and a radius cliff; while the radius distribution for planets in binaries shows no clear radius valley and much lower occurrence than for the single stars. In addition, the difference in shape between the single and binary star planet occurrence distributions is $\sim 4.3 \sigma$ significant via a Kolmogorov-Smirnov test. This result confirms the observational result that planets in binaries are intrinsically less common than those in single stars. The distribution also does not show evidence for a radius valley, supporting the results from both \citet{Sullivan2023} and \citet{Sullivan2024b}.

In \citet{Sullivan2024b}, the authors' explanation for the unimodal radius distribution and overabundance of super-Earths relative to sub-Neptunes was that the shorter disk lifetimes in binaries \citep[e.g.,][]{Cieza2009, Harris2012, Kraus2012} lead to formation of fewer sub-Neptunes, which suppresses the appearance of the radius valley. However, Figure \ref{fig:occ_rgap} tells a slightly different story. In occurrence space, the relative number of super-Earths to sub-Neptunes is 1.81$\pm$0.27 times more super-Earths than sub-Neptunes, which is less than the factor of 2.47$\pm$0.31 excess in the observed planet distribution \citep{Sullivan2024b}. The difference between the two ratios is small, and likely because of a low completeness limit that was not accounted for in the previous work: even though there are planets in that regime (accounted for in the \citet{Sullivan2024b} work), they are not in a high enough completeness regime to be counted for the occurrence calculation.

Although the relative number of super-Earths and sub-Neptunes may differ in occurrence space compared to observed space, one feature that remains the same is the lack of a radius valley. Instead, the distribution is flat within error, and the lowest point of the radius valley for single stars is one of the locations where the two distributions agree in occurrence. There is also no clear evidence of a radius cliff, although this may be because of the lower overall occurrence, which again agrees with the single-star distribution around 1 Neptune radius or $4 \,{\rm R_{\oplus}}$, and becomes equally numerous as the planets around single stars at large radii (see Figure \ref{fig:ratio_map}). 

Another possible reason for the lack of a radius valley is the relatively small planet sample in each catalog ($\sim$ 30 planets), which may not be sufficient to resolve the valley. However, while each catalog contains few planets, they all have planets that cover the small planet radius regime, and the total catalog of $\sim$ 5000 stars includes enough planets ($\sim$ 100) that we expect we could resolve the radius valley. However, our error bars on each occurrence bin are relatively large because of the small number of stars and planets. A larger sample and smaller error bars might resolve a radius valley.

The lack of an apparent observed radius valley for planets in binaries corroborates other results in the literature but remains tentative because of the small sample size, warranting further investigation. Assuming that the radius valley is missing, we can hypothesize about possible causes for the systematically low occurrence rate for planets in binaries. The overall low occurrence rate for binaries is likely a result of shorter disk lifetimes and lower disk masses suppressing planet formation, as well as a dynamically hotter system during the early stages of the planet's evolution potentially causing planet ejection from the system \citep[e.g.,][]{JangCondell2015, Kraus2016, Sullivan2024b}. The lack of a radius valley is more difficult to explain, because it is expected to be a fundamental feature of planet populations. Recent work has suggested that planets in the radius valley have higher eccentricities than those outside \citep{Gilbert2025}. The origin of the elevated eccentricities, as well as the population of the planets inside the radius valley, may be explained as the result of an epoch of giant impacts \citep{Ho2024, Shibata2025}. If small-separation binaries (separations on the order of 100\,au) are dynamically hotter than their single-star counterparts because of the stellar companion, dynamical disruption of planets and increased rates of giant impacts may occur, filling in the radius valley. However, more theoretical work is necessary to support or refute this possibility. 

\subsection{Conclusions}\label{sec:conclusions}
We assembled and characterized a large sample of high-likelihood binaries in the KTC using multiplicity indicators from Gaia, \citet{Berger2018}, and \citet{Berger2020stars}, as well as stellar characterization from \citet{Berger2020stars} and binary properties inferred using MOLUSC \citep{Wood2021}. We used our new stellar catalog and adjustments to the standard occurrence rate methodologies for single stars \citep[e.g.,][]{Bryson2020,Dattilo2023} to calculate the first occurrence rates for planets in binary star systems. We also investigated the small planet radius distribution in binary systems in occurrence space, and compared to results from \citet{Sullivan2023} and \citet{Sullivan2024b}. 

We found the following:
\begin{itemize}
    \item The photometric noise properties of binary stars in the KTC are comparable to those of single stars (Figure \ref{fig:cdpp_slope}).
    \item The integrated occurrence rate for planets in binaries over 1-10 $R_{\oplus}$ and 1-100 d is one-half that of single stars, to 3.8$\sigma$ significance.
    \item The integrated occurrence rate over a smaller period-radius range, 1-4 $R_{\oplus}$ and 1-50 d, with higher completeness and more planet detections, is much lower, with 58\% fewer planets in binaries to 11.4$\sigma$ significance.
    \item The radius distribution for small planets in binary systems, when marginalized over 1-50\,d orbital periods, is distinct from the radius distribution for planets in single stars to 4.3$\sigma$, and does not show evidence for a radius cliff or a radius valley. This difference indicates distinct planet formation and survival mechanisms for planets in small-separation binary systems compared to single stars, likely because of the impact of the secondary star on the mass and lifetime of the protoplanetary disk, as well as additional dynamical disruption from the secondary star.
\end{itemize}

Our results are estimates of planet occurrence rates in binary systems, calculated in a probabilistic way rather than using a sample of definitively known binary systems. Future work should explore occurrence for planets in small-separation binaries in other surveys, ideally where small-separation binaries can be more easily identified and characterized. For example, nearby planets found by the Transiting Exoplanet Survey Satellite (TESS; \citealt{Ricker2015}) are typically around nearby stars, where small-physical-separation binaries will have relatively larger angular separations and are thus easier to detect. Gaia DR4, expected to be released in 2026, will include a variety of different samples of binary stars, mostly detected using astrometry, which will provide substantial additional stellar information that can be used to refine analyses like the effort presented in this paper. Finally, the PLAnetary Transits and Oscillations (PLATO) mission \citep{Rauer2014} is scheduled to launch in 2026 and begin long-time-baseline observations of multiple stellar fields to find transiting exoplanets. One of the major goals of PLATO is to include characterization for its stellar catalog which should lead to a large sample of binary stars, including planet hosts, with uniform stellar properties that will lend itself to additional occurrence rate studies. 

Beyond observational efforts, theoretical studies would be beneficial to understand the origins of the difference in the structure of the planet occurrence between planets in binaries versus those in single stars. The radius valley and radius cliff are both expected to be fundamental properties of the planet population for single stars, produced by underlying physics that should be universal for planet formation. The lack of these features for planets in binary star systems suggests that additional, complex physics is occurring to remove or alter planets forming and evolving in binary star systems. Planets in binary stars thus also offer an exciting opportunity to explore the outcomes of planet formation in extreme environments.

\software{astropy \citep{astropy2013, astropy2018, astropy2022}, corner \citep{Foreman-Mackey2016}, emcee \citep{Foreman-Mackey2013}, matplotlib \citep{Hunter2007}, numpy \citep{Harris2020}, pyphot \citep{pyphot}, scipy \citep{Virtanen2020}, scikit-learn \citep{scikit-learn}}
\facilities{Kepler \citep{Borucki2010}, Gaia \citep{Gaia2016}, Exoplanet Archive \citep{exoarchive}}

\begin{acknowledgments}
We thank the referee for their thoughtful comments, which improved this manuscript. NMB acknowledges support from NASA’S Interdisciplinary Consortia for Astrobiology Research (NNH19ZDA001N-ICAR) under award number 80NSSC21K0597. AD gratefully acknowledges support from the Heising-Simons Foundation through grant 2021-3197. This research has made use of the VizieR catalogue access tool, CDS, Strasbourg Astronomical Observatory, France (DOI : 10.26093/cds/vizier; \citealt{vizier}). This work has made use of data from the European Space Agency (ESA) mission {\it Gaia} (\url{https://www.cosmos.esa.int/gaia}), processed by the {\it Gaia} Data Processing and Analysis Consortium (DPAC, \url{https://www.cosmos.esa.int/web/gaia/dpac/consortium}). Funding for the DPAC has been provided by national institutions, in particular the institutions participating in the {\it Gaia} Multilateral Agreement. This research has made use of the NASA Exoplanet Archive, which is operated by the California Institute of Technology, under contract with the National Aeronautics and Space Administration under the Exoplanet Exploration Program, including the following Tables: \citet{Planet_catalog_DOI}, \citet{Kepler_completeness_DOI}.
\end{acknowledgments}

\bibliographystyle{aasjournalv7}
\bibliography{bib}

\begin{thebibliography}{}
\expandafter\ifx\csname natexlab\endcsname\relax\def\natexlab#1{#1}\fi
\providecommand{\url}[1]{\href{#1}{#1}}
\providecommand{\dodoi}[1]{doi:~\href{http://doi.org/#1}{\nolinkurl{#1}}}
\providecommand{\doeprint}[1]{\href{http://ascl.net/#1}{\nolinkurl{http://ascl.net/#1}}}
\providecommand{\doarXiv}[1]{\href{https://arxiv.org/abs/#1}{\nolinkurl{https://arxiv.org/abs/#1}}}

\bibitem[{ {Astropy Collaboration} {et~al.}(2013){Astropy Collaboration},
  {Robitaille}, {Tollerud}, {Greenfield}, {Droettboom}, {Bray}, {Aldcroft},
  {Davis}, {Ginsburg}, {Price-Whelan}, {Kerzendorf}, {Conley}, {Crighton},
  {Barbary}, {Muna}, {Ferguson}, {Grollier}, {Parikh}, {Nair}, {Unther},
  {Deil}, {Woillez}, {Conseil}, {Kramer}, {Turner}, {Singer}, {Fox}, {Weaver},
  {Zabalza}, {Edwards}, {Azalee Bostroem}, {Burke}, {Casey}, {Crawford},
  {Dencheva}, {Ely}, {Jenness}, {Labrie}, {Lim}, {Pierfederici}, {Pontzen},
  {Ptak}, {Refsdal}, {Servillat}, \& {Streicher}}]{astropy2013}
{Astropy Collaboration}, {Robitaille}, T.~P., {Tollerud}, E.~J., {et~al.} 2013,
  \bibinfo{title}{{Astropy: A community Python package for astronomy},} \aap,
  558, A33, \dodoi{10.1051/0004-6361/201322068}

\bibitem[{ {Astropy Collaboration} {et~al.}(2018){Astropy Collaboration},
  {Price-Whelan}, {Sip{\H{o}}cz}, {G{\"u}nther}, {Lim}, {Crawford}, {Conseil},
  {Shupe}, {Craig}, {Dencheva}, {Ginsburg}, {Vand erPlas}, {Bradley},
  {P{\'e}rez-Su{\'a}rez}, {de Val-Borro}, {Aldcroft}, {Cruz}, {Robitaille},
  {Tollerud}, {Ardelean}, {Babej}, {Bach}, {Bachetti}, {Bakanov}, {Bamford},
  {Barentsen}, {Barmby}, {Baumbach}, {Berry}, {Biscani}, {Boquien}, {Bostroem},
  {Bouma}, {Brammer}, {Bray}, {Breytenbach}, {Buddelmeijer}, {Burke},
  {Calderone}, {Cano Rodr{\'\i}guez}, {Cara}, {Cardoso}, {Cheedella}, {Copin},
  {Corrales}, {Crichton}, {D'Avella}, {Deil}, {Depagne}, {Dietrich}, {Donath},
  {Droettboom}, {Earl}, {Erben}, {Fabbro}, {Ferreira}, {Finethy}, {Fox},
  {Garrison}, {Gibbons}, {Goldstein}, {Gommers}, {Greco}, {Greenfield},
  {Groener}, {Grollier}, {Hagen}, {Hirst}, {Homeier}, {Horton}, {Hosseinzadeh},
  {Hu}, {Hunkeler}, {Ivezi{\'c}}, {Jain}, {Jenness}, {Kanarek}, {Kendrew},
  {Kern}, {Kerzendorf}, {Khvalko}, {King}, {Kirkby}, {Kulkarni}, {Kumar},
  {Lee}, {Lenz}, {Littlefair}, {Ma}, {Macleod}, {Mastropietro}, {McCully},
  {Montagnac}, {Morris}, {Mueller}, {Mumford}, {Muna}, {Murphy}, {Nelson},
  {Nguyen}, {Ninan}, {N{\"o}the}, {Ogaz}, {Oh}, {Parejko}, {Parley}, {Pascual},
  {Patil}, {Patil}, {Plunkett}, {Prochaska}, {Rastogi}, {Reddy Janga},
  {Sabater}, {Sakurikar}, {Seifert}, {Sherbert}, {Sherwood-Taylor}, {Shih},
  {Sick}, {Silbiger}, {Singanamalla}, {Singer}, {Sladen}, {Sooley},
  {Sornarajah}, {Streicher}, {Teuben}, {Thomas}, {Tremblay}, {Turner},
  {Terr{\'o}n}, {van Kerkwijk}, {de la Vega}, {Watkins}, {Weaver}, {Whitmore},
  {Woillez}, {Zabalza}, \& {Astropy Contributors}}]{astropy2018}
{Astropy Collaboration}, {Price-Whelan}, A.~M., {Sip{\H{o}}cz}, B.~M., {et~al.}
  2018, \bibinfo{title}{{The Astropy Project: Building an Open-science Project
  and Status of the v2.0 Core Package},} \aj, 156, 123,
  \dodoi{10.3847/1538-3881/aabc4f}

\bibitem[{ {Astropy Collaboration} {et~al.}(2022){Astropy Collaboration},
  {Price-Whelan}, {Lim}, {Earl}, {Starkman}, {Bradley}, {Shupe}, {Patil},
  {Corrales}, {Brasseur}, {N{\"o}the}, {Donath}, {Tollerud}, {Morris},
  {Ginsburg}, {Vaher}, {Weaver}, {Tocknell}, {Jamieson}, {van Kerkwijk},
  {Robitaille}, {Merry}, {Bachetti}, {G{\"u}nther}, {Aldcroft},
  {Alvarado-Montes}, {Archibald}, {B{\'o}di}, {Bapat}, {Barentsen},
  {Baz{\'a}n}, {Biswas}, {Boquien}, {Burke}, {Cara}, {Cara}, {Conroy},
  {Conseil}, {Craig}, {Cross}, {Cruz}, {D'Eugenio}, {Dencheva}, {Devillepoix},
  {Dietrich}, {Eigenbrot}, {Erben}, {Ferreira}, {Foreman-Mackey}, {Fox},
  {Freij}, {Garg}, {Geda}, {Glattly}, {Gondhalekar}, {Gordon}, {Grant},
  {Greenfield}, {Groener}, {Guest}, {Gurovich}, {Handberg}, {Hart},
  {Hatfield-Dodds}, {Homeier}, {Hosseinzadeh}, {Jenness}, {Jones}, {Joseph},
  {Kalmbach}, {Karamehmetoglu}, {Ka{\l}uszy{\'n}ski}, {Kelley}, {Kern},
  {Kerzendorf}, {Koch}, {Kulumani}, {Lee}, {Ly}, {Ma}, {MacBride}, {Maljaars},
  {Muna}, {Murphy}, {Norman}, {O'Steen}, {Oman}, {Pacifici}, {Pascual},
  {Pascual-Granado}, {Patil}, {Perren}, {Pickering}, {Rastogi}, {Roulston},
  {Ryan}, {Rykoff}, {Sabater}, {Sakurikar}, {Salgado}, {Sanghi}, {Saunders},
  {Savchenko}, {Schwardt}, {Seifert-Eckert}, {Shih}, {Jain}, {Shukla}, {Sick},
  {Simpson}, {Singanamalla}, {Singer}, {Singhal}, {Sinha}, {Sip{\H{o}}cz},
  {Spitler}, {Stansby}, {Streicher}, {{\v{S}}umak}, {Swinbank}, {Taranu},
  {Tewary}, {Tremblay}, {Val-Borro}, {Van Kooten}, {Vasovi{\'c}}, {Verma}, {de
  Miranda Cardoso}, {Williams}, {Wilson}, {Winkel}, {Wood-Vasey}, {Xue},
  {Yoachim}, {Zhang}, {Zonca}, \& {Astropy Project Contributors}}]{astropy2022}
{Astropy Collaboration}, {Price-Whelan}, A.~M., {Lim}, P.~L., {et~al.} 2022,
  \bibinfo{title}{{The Astropy Project: Sustaining and Growing a
  Community-oriented Open-source Project and the Latest Major Release (v5.0) of
  the Core Package},} \apj, 935, 167, \dodoi{10.3847/1538-4357/ac7c74}

\bibitem[{N.~M. {Batalha} {et~al.}(2010){Batalha}, {Borucki}, {Koch}, {Bryson},
  {Haas}, {Brown}, {Caldwell}, {Hall}, {Gilliland}, {Latham}, {Meibom}, \&
  {Monet}}]{Batalha2010}
{Batalha}, N.~M., {Borucki}, W.~J., {Koch}, D.~G., {et~al.} 2010,
  \bibinfo{title}{{Selection, Prioritization, and Characteristics of Kepler
  Target Stars},} \apjl, 713, L109, \dodoi{10.1088/2041-8205/713/2/L109}

\bibitem[{T.~A. {Berger} {et~al.}(2018){Berger}, {Huber}, {Gaidos}, \& {van
  Saders}}]{Berger2018}
{Berger}, T.~A., {Huber}, D., {Gaidos}, E., \& {van Saders}, J.~L. 2018,
  \bibinfo{title}{{Revised Radii of Kepler Stars and Planets Using Gaia Data
  Release 2},} \apj, 866, 99, \dodoi{10.3847/1538-4357/aada83}

\bibitem[{T.~A. {Berger} {et~al.}(2020){Berger}, {Huber}, {van Saders},
  {Gaidos}, {Tayar}, \& {Kraus}}]{Berger2020stars}
{Berger}, T.~A., {Huber}, D., {van Saders}, J.~L., {et~al.} 2020,
  \bibinfo{title}{{The Gaia-Kepler Stellar Properties Catalog. I. Homogeneous
  Fundamental Properties for 186,301 Kepler Stars},} \aj, 159, 280,
  \dodoi{10.3847/1538-3881/159/6/280}

\bibitem[{G.~J. {Bergsten} {et~al.}(2022){Bergsten}, {Pascucci}, {Mulders},
  {Fernandes}, \& {Koskinen}}]{Bergsten2022}
{Bergsten}, G.~J., {Pascucci}, I., {Mulders}, G.~D., {Fernandes}, R.~B., \&
  {Koskinen}, T.~T. 2022, \bibinfo{title}{{The Demographics of Kepler's Earths
  and Super-Earths into the Habitable Zone},} \aj, 164, 190,
  \dodoi{10.3847/1538-3881/ac8fea}

\bibitem[{W.~J. {Borucki} {et~al.}(2010){Borucki}, {Koch}, {Basri}, {Batalha},
  {Brown}, {Caldwell}, {Caldwell}, {Christensen-Dalsgaard}, {Cochran},
  {DeVore}, {Dunham}, {Dupree}, {Gautier}, {Geary}, {Gilliland}, {Gould},
  {Howell}, {Jenkins}, {Kondo}, {Latham}, {Marcy}, {Meibom}, {Kjeldsen},
  {Lissauer}, {Monet}, {Morrison}, {Sasselov}, {Tarter}, {Boss}, {Brownlee},
  {Owen}, {Buzasi}, {Charbonneau}, {Doyle}, {Fortney}, {Ford}, {Holman},
  {Seager}, {Steffen}, {Welsh}, {Rowe}, {Anderson}, {Buchhave}, {Ciardi},
  {Walkowicz}, {Sherry}, {Horch}, {Isaacson}, {Everett}, {Fischer}, {Torres},
  {Johnson}, {Endl}, {MacQueen}, {Bryson}, {Dotson}, {Haas}, {Kolodziejczak},
  {Van Cleve}, {Chandrasekaran}, {Twicken}, {Quintana}, {Clarke}, {Allen},
  {Li}, {Wu}, {Tenenbaum}, {Verner}, {Bruhweiler}, {Barnes}, \&
  {Prsa}}]{Borucki2010}
{Borucki}, W.~J., {Koch}, D., {Basri}, G., {et~al.} 2010,
  \bibinfo{title}{{Kepler Planet-Detection Mission: Introduction and First
  Results},} Science, 327, 977, \dodoi{10.1126/science.1185402}

\bibitem[{T.~M. {Brown} {et~al.}(2011){Brown}, {Latham}, {Everett}, \&
  {Esquerdo}}]{Brown2011}
{Brown}, T.~M., {Latham}, D.~W., {Everett}, M.~E., \& {Esquerdo}, G.~A. 2011,
  \bibinfo{title}{{Kepler Input Catalog: Photometric Calibration and Stellar
  Classification},} \aj, 142, 112, \dodoi{10.1088/0004-6256/142/4/112}

\bibitem[{S. {Bryson} {et~al.}(2020{\natexlab{a}}){Bryson}, {Coughlin},
  {Batalha}, {Berger}, {Huber}, {Burke}, {Dotson}, \& {Mullally}}]{Bryson2020}
{Bryson}, S., {Coughlin}, J., {Batalha}, N.~M., {et~al.} 2020{\natexlab{a}},
  \bibinfo{title}{{A Probabilistic Approach to Kepler Completeness and
  Reliability for Exoplanet Occurrence Rates},} \aj, 159, 279,
  \dodoi{10.3847/1538-3881/ab8a30}

\bibitem[{S. {Bryson} {et~al.}(2020{\natexlab{b}}){Bryson}, {Coughlin},
  {Kunimoto}, \& {Mullally}}]{Bryson2020b}
{Bryson}, S., {Coughlin}, J.~L., {Kunimoto}, M., \& {Mullally}, S.~E.
  2020{\natexlab{b}}, \bibinfo{title}{{Reliability Correction is Key for Robust
  Kepler Occurrence Rates},} \aj, 160, 200, \dodoi{10.3847/1538-3881/abb316}

\bibitem[{S. {Bryson} {et~al.}(2021){Bryson}, {Kunimoto}, {Kopparapu},
  {Coughlin}, {Borucki}, {Koch}, {Aguirre}, {Allen}, {Barentsen}, {Batalha},
  {Berger}, {Boss}, {Buchhave}, {Burke}, {Caldwell}, {Campbell}, {Catanzarite},
  {Chandrasekaran}, {Chaplin}, {Christiansen}, {Christensen-Dalsgaard},
  {Ciardi}, {Clarke}, {Cochran}, {Dotson}, {Doyle}, {Duarte}, {Dunham},
  {Dupree}, {Endl}, {Fanson}, {Ford}, {Fujieh}, {Gautier}, {Geary},
  {Gilliland}, {Girouard}, {Gould}, {Haas}, {Henze}, {Holman}, {Howard},
  {Howell}, {Huber}, {Hunter}, {Jenkins}, {Kjeldsen}, {Kolodziejczak},
  {Larson}, {Latham}, {Li}, {Mathur}, {Meibom}, {Middour}, {Morris}, {Morton},
  {Mullally}, {Mullally}, {Pletcher}, {Prsa}, {Quinn}, {Quintana}, {Ragozzine},
  {Ramirez}, {Sanderfer}, {Sasselov}, {Seader}, {Shabram}, {Shporer}, {Smith},
  {Steffen}, {Still}, {Torres}, {Troeltzsch}, {Twicken}, {Uddin}, {Van Cleve},
  {Voss}, {Weiss}, {Welsh}, {Wohler}, \& {Zamudio}}]{Bryson2021}
{Bryson}, S., {Kunimoto}, M., {Kopparapu}, R.~K., {et~al.} 2021,
  \bibinfo{title}{{The Occurrence of Rocky Habitable-zone Planets around
  Solar-like Stars from Kepler Data},} \aj, 161, 36,
  \dodoi{10.3847/1538-3881/abc418}

\bibitem[{C.~J. {Burke} \& J. {Catanzarite}(2017{\natexlab{a}}){Burke} \&
  {Catanzarite}}]{Burke2017}
{Burke}, C.~J., \& {Catanzarite}, J. 2017{\natexlab{a}},
  \bibinfo{title}{{Planet Detection Metrics: Per-Target Detection Contours for
  Data Release 25},}, Kepler Science Document KSCI-19111-002, id. 19. Edited by
  Michael R. Haas and Natalie M. Batalha

\bibitem[{C.~J. {Burke} \& J. {Catanzarite}(2017{\natexlab{b}}){Burke} \&
  {Catanzarite}}]{Burke2017onesigma}
{Burke}, C.~J., \& {Catanzarite}, J. 2017{\natexlab{b}},
  \bibinfo{title}{{Planet Detection Metrics: Window and One-Sigma Depth
  Functions for Data Release 25},}, Kepler Science Document KSCI-19101-002, id.
  14. Edited by Michael R. Haas and Natalie M. Batalha

\bibitem[{C.~J. {Burke} {et~al.}(2015){Burke}, {Christiansen}, {Mullally},
  {Seader}, {Huber}, {Rowe}, {Coughlin}, {Thompson}, {Catanzarite}, {Clarke},
  {Morton}, {Caldwell}, {Bryson}, {Haas}, {Batalha}, {Jenkins}, {Tenenbaum},
  {Twicken}, {Li}, {Quintana}, {Barclay}, {Henze}, {Borucki}, {Howell}, \&
  {Still}}]{Burke2015}
{Burke}, C.~J., {Christiansen}, J.~L., {Mullally}, F., {et~al.} 2015,
  \bibinfo{title}{{Terrestrial Planet Occurrence Rates for the Kepler GK Dwarf
  Sample},} \apj, 809, 8, \dodoi{10.1088/0004-637X/809/1/8}

\bibitem[{J. {Choi} {et~al.}(2016){Choi}, {Dotter}, {Conroy}, {Cantiello},
  {Paxton}, \& {Johnson}}]{Choi2016}
{Choi}, J., {Dotter}, A., {Conroy}, C., {et~al.} 2016, \bibinfo{title}{{Mesa
  Isochrones and Stellar Tracks (MIST). I. Solar-scaled Models},} \apj, 823,
  102, \dodoi{10.3847/0004-637X/823/2/102}

\bibitem[{J.~L. {Christiansen}(2017){Christiansen}}]{Christiansen2017}
{Christiansen}, J.~L. 2017, \bibinfo{title}{{Planet Detection Metrics:
  Pixel-Level Transit Injection Tests of Pipeline Detection Efficiency for Data
  Release 25},}, Kepler Science Document KSCI-19110-001, id. 18. Edited by
  Michael R. Haas and Natalie M. Batalha

\bibitem[{J.~L. {Christiansen} {et~al.}(2025){Christiansen}, {McElroy},
  {Harbut}, {Ciardi}, {Crane}, {Good}, {Hardegree-Ullman}, {Kesseli}, {Lund},
  {Lynn}, {Muthiar}, {Nilsson}, {Oluyide}, {Papin}, {Rivera}, {Swain},
  {Susemiehl}, {Tam}, {van Eyken}, \& {Beichman}}]{exoarchive}
{Christiansen}, J.~L., {McElroy}, D.~L., {Harbut}, M., {et~al.} 2025,
  \bibinfo{title}{{The NASA Exoplanet Archive and Exoplanet Follow-up Observing
  Program: Data, Tools, and Usage},} arXiv e-prints, arXiv:2506.03299.
\newblock \doarXiv{2506.03299}

\bibitem[{D.~R. {Ciardi} {et~al.}(2015){Ciardi}, {Beichman}, {Horch}, \&
  {Howell}}]{Ciardi2015}
{Ciardi}, D.~R., {Beichman}, C.~A., {Horch}, E.~P., \& {Howell}, S.~B. 2015,
  \bibinfo{title}{{Understanding the Effects of Stellar Multiplicity on the
  Derived Planet Radii from Transit Surveys: Implications for Kepler, K2, and
  TESS},} \apj, 805, 16, \dodoi{10.1088/0004-637X/805/1/16}

\bibitem[{L.~A. {Cieza} {et~al.}(2009){Cieza}, {Padgett}, {Allen}, {McCabe},
  {Brooke}, {Carey}, {Chapman}, {Fukagawa}, {Huard}, {Noriga-Crespo},
  {Peterson}, \& {Rebull}}]{Cieza2009}
{Cieza}, L.~A., {Padgett}, D.~L., {Allen}, L.~E., {et~al.} 2009,
  \bibinfo{title}{{Primordial Circumstellar Disks in Binary Systems: Evidence
  for Reduced Lifetimes},} \apjl, 696, L84, \dodoi{10.1088/0004-637X/696/1/L84}

\bibitem[{C.~A. {Clark} {et~al.}(2024){Clark}, {van Belle}, {Horch}, {Lund},
  {Ciardi}, {von Braun}, {Winters}, {Everett}, {Hartman}, \&
  {Llama}}]{Clark2024}
{Clark}, C.~A., {van Belle}, G.~T., {Horch}, E.~P., {et~al.} 2024,
  \bibinfo{title}{{The POKEMON Speckle Survey of Nearby M Dwarfs. II.
  Observations of 1125 Targets},} \aj, 167, 56,
  \dodoi{10.3847/1538-3881/ad0bfd}

\bibitem[{J.~L. {Coughlin} {et~al.}(2016){Coughlin}, {Mullally}, {Thompson},
  {Rowe}, {Burke}, {Latham}, {Batalha}, {Ofir}, {Quarles}, {Henze}, {Wolfgang},
  {Caldwell}, {Bryson}, {Shporer}, {Catanzarite}, {Akeson}, {Barclay},
  {Borucki}, {Boyajian}, {Campbell}, {Christiansen}, {Girouard}, {Haas},
  {Howell}, {Huber}, {Jenkins}, {Li}, {Patil-Sabale}, {Quintana}, {Ramirez},
  {Seader}, {Smith}, {Tenenbaum}, {Twicken}, \& {Zamudio}}]{Coughlin2016}
{Coughlin}, J.~L., {Mullally}, F., {Thompson}, S.~E., {et~al.} 2016,
  \bibinfo{title}{{Planetary Candidates Observed by Kepler. VII. The First
  Fully Uniform Catalog Based on the Entire 48-month Data Set (Q1-Q17 DR24)},}
  \apjs, 224, 12, \dodoi{10.3847/0067-0049/224/1/12}

\bibitem[{A. {Dattilo} \& N.~M. {Batalha}(2024){Dattilo} \&
  {Batalha}}]{Dattilo2024}
{Dattilo}, A., \& {Batalha}, N.~M. 2024, \bibinfo{title}{{A Unified Treatment
  of Kepler Occurrence to Trace Planet Evolution. II. The Radius Cliff Formed
  by Atmospheric Escape},} \aj, 167, 288, \dodoi{10.3847/1538-3881/ad434c}

\bibitem[{A. {Dattilo} {et~al.}(2023){Dattilo}, {Batalha}, \&
  {Bryson}}]{Dattilo2023}
{Dattilo}, A., {Batalha}, N.~M., \& {Bryson}, S. 2023, \bibinfo{title}{{A
  Unified Treatment of Kepler Occurrence to Trace Planet Evolution. I.
  Methodology},} \aj, 166, 122, \dodoi{10.3847/1538-3881/acebc8}

\bibitem[{J.~M. {Dodd} {et~al.}(2024){Dodd}, {Oudmaijer}, {Radley}, {Vioque},
  \& {Frost}}]{Dodd2024}
{Dodd}, J.~M., {Oudmaijer}, R.~D., {Radley}, I.~C., {Vioque}, M., \& {Frost},
  A.~J. 2024, \bibinfo{title}{{Gaia uncovers difference in B and Be star
  binarity at small scales: evidence for mass transfer causing the Be
  phenomenon},} \mnras, 527, 3076, \dodoi{10.1093/mnras/stad3105}

\bibitem[{A. {Dotter}(2016){Dotter}}]{Dotter2016}
{Dotter}, A. 2016, \bibinfo{title}{{MESA Isochrones and Stellar Tracks (MIST)
  0: Methods for the Construction of Stellar Isochrones},} \apjs, 222, 8,
  \dodoi{10.3847/0067-0049/222/1/8}

\bibitem[{D. Foreman-Mackey(2016)Foreman-Mackey}]{Foreman-Mackey2016}
Foreman-Mackey, D. 2016, \bibinfo{title}{corner.py: Scatterplot matrices in
  Python,} Journal of Open Source Software, 1, 24, \dodoi{10.21105/joss.00024}

\bibitem[{D. {Foreman-Mackey} {et~al.}(2013){Foreman-Mackey}, {Hogg}, {Lang},
  \& {Goodman}}]{Foreman-Mackey2013}
{Foreman-Mackey}, D., {Hogg}, D.~W., {Lang}, D., \& {Goodman}, J. 2013,
  \bibinfo{title}{{emcee: The MCMC Hammer},} \pasp, 125, 306,
  \dodoi{10.1086/670067}

\bibitem[{D. {Foreman-Mackey} {et~al.}(2014){Foreman-Mackey}, {Hogg}, \&
  {Morton}}]{Foreman-Mackey2014}
{Foreman-Mackey}, D., {Hogg}, D.~W., \& {Morton}, T.~D. 2014,
  \bibinfo{title}{{Exoplanet Population Inference and the Abundance of Earth
  Analogs from Noisy, Incomplete Catalogs},} \apj, 795, 64,
  \dodoi{10.1088/0004-637X/795/1/64}

\bibitem[{M. Fouesneau(2022)Fouesneau}]{pyphot}
Fouesneau, M. 2022, \bibinfo{title}{pyphot,} Zenodo,
  \dodoi{10.5281/ZENODO.7016775}

\bibitem[{B.~J. {Fulton} {et~al.}(2017){Fulton}, {Petigura}, {Howard},
  {Isaacson}, {Marcy}, {Cargile}, {Hebb}, {Weiss}, {Johnson}, {Morton},
  {Sinukoff}, {Crossfield}, \& {Hirsch}}]{Fulton2017}
{Fulton}, B.~J., {Petigura}, E.~A., {Howard}, A.~W., {et~al.} 2017,
  \bibinfo{title}{{The California-Kepler Survey. III. A Gap in the Radius
  Distribution of Small Planets},} \aj, 154, 109,
  \dodoi{10.3847/1538-3881/aa80eb}

\bibitem[{E. {Furlan} \& S.~B. {Howell}(2020){Furlan} \& {Howell}}]{Furlan2020}
{Furlan}, E., \& {Howell}, S.~B. 2020, \bibinfo{title}{{Unresolved Binary
  Exoplanet Host Stars Fit as Single Stars: Effects on the Stellar
  Parameters},} \apj, 898, 47, \dodoi{10.3847/1538-4357/ab9c9c}

\bibitem[{E. {Furlan} {et~al.}(2017){Furlan}, {Ciardi}, {Everett}, {Saylors},
  {Teske}, {Horch}, {Howell}, {van Belle}, {Hirsch}, {Gautier}, {Adams},
  {Barrado}, {Cartier}, {Dressing}, {Dupree}, {Gilliland}, {Lillo-Box},
  {Lucas}, \& {Wang}}]{Furlan2017}
{Furlan}, E., {Ciardi}, D.~R., {Everett}, M.~E., {et~al.} 2017,
  \bibinfo{title}{{The Kepler Follow-up Observation Program. I. A Catalog of
  Companions to Kepler Stars from High-Resolution Imaging},} \aj, 153, 71,
  \dodoi{10.3847/1538-3881/153/2/71}

\bibitem[{ {Gaia Collaboration} {et~al.}(2016){Gaia Collaboration}, {Prusti},
  {de Bruijne}, {Brown}, {Vallenari}, {Babusiaux}, {Bailer-Jones}, {Bastian},
  {Biermann}, {Evans}, {Eyer}, {Jansen}, {Jordi}, {Klioner}, {Lammers},
  {Lindegren}, {Luri}, {Mignard}, {Milligan}, {Panem}, {Poinsignon},
  {Pourbaix}, {Randich}, {Sarri}, {Sartoretti}, {Siddiqui}, {Soubiran},
  {Valette}, {van Leeuwen}, {Walton}, {Aerts}, {Arenou}, {Cropper}, {Drimmel},
  {H{\o}g}, {Katz}, {Lattanzi}, {O'Mullane}, {Grebel}, {Holland}, {Huc},
  {Passot}, {Bramante}, {Cacciari}, {Casta{\~n}eda}, {Chaoul}, {Cheek}, {De
  Angeli}, {Fabricius}, {Guerra}, {Hern{\'a}ndez}, {Jean-Antoine-Piccolo},
  {Masana}, {Messineo}, {Mowlavi}, {Nienartowicz}, {Ord{\'o}{\~n}ez-Blanco},
  {Panuzzo}, {Portell}, {Richards}, {Riello}, {Seabroke}, {Tanga},
  {Th{\'e}venin}, {Torra}, {Els}, {Gracia-Abril}, {Comoretto},
  {Garcia-Reinaldos}, {Lock}, {Mercier}, {Altmann}, {Andrae}, {Astraatmadja},
  {Bellas-Velidis}, {Benson}, {Berthier}, {Blomme}, {Busso}, {Carry},
  {Cellino}, {Clementini}, {Cowell}, {Creevey}, {Cuypers}, {Davidson}, {De
  Ridder}, {de Torres}, {Delchambre}, {Dell'Oro}, {Ducourant}, {Fr{\'e}mat},
  {Garc{\'\i}a-Torres}, {Gosset}, {Halbwachs}, {Hambly}, {Harrison}, {Hauser},
  {Hestroffer}, {Hodgkin}, {Huckle}, {Hutton}, {Jasniewicz}, {Jordan},
  {Kontizas}, {Korn}, {Lanzafame}, {Manteiga}, {Moitinho}, {Muinonen},
  {Osinde}, {Pancino}, {Pauwels}, {Petit}, {Recio-Blanco}, {Robin}, {Sarro},
  {Siopis}, {Smith}, {Smith}, {Sozzetti}, {Thuillot}, {van Reeven}, {Viala},
  {Abbas}, {Abreu Aramburu}, {Accart}, {Aguado}, {Allan}, {Allasia},
  {Altavilla}, {{\'A}lvarez}, {Alves}, {Anderson}, {Andrei}, {Anglada Varela},
  {Antiche}, {Antoja}, {Ant{\'o}n}, {Arcay}, {Atzei}, {Ayache}, {Bach},
  {Baker}, {Balaguer-N{\'u}{\~n}ez}, {Barache}, {Barata}, {Barbier}, {Barblan},
  {Baroni}, {Barrado y Navascu{\'e}s}, {Barros}, {Barstow}, {Becciani},
  {Bellazzini}, {Bellei}, {Bello Garc{\'\i}a}, {Belokurov}, {Bendjoya},
  {Berihuete}, {Bianchi}, {Bienaym{\'e}}, {Billebaud}, {Blagorodnova},
  {Blanco-Cuaresma}, {Boch}, {Bombrun}, {Borrachero}, {Bouquillon}, {Bourda},
  {Bouy}, {Bragaglia}, {Breddels}, {Brouillet}, {Br{\"u}semeister},
  {Bucciarelli}, {Budnik}, {Burgess}, {Burgon}, {Burlacu}, {Busonero}, {Buzzi},
  {Caffau}, {Cambras}, {Campbell}, {Cancelliere}, {Cantat-Gaudin}, {Carlucci},
  {Carrasco}, {Castellani}, {Charlot}, {Charnas}, {Charvet}, {Chassat},
  {Chiavassa}, {Clotet}, {Cocozza}, {Collins}, {Collins}, {Costigan}, {Crifo},
  {Cross}, {Crosta}, {Crowley}, {Dafonte}, {Damerdji}, {Dapergolas}, {David},
  {David}, {De Cat}, {de Felice}, {de Laverny}, {De Luise}, {De March}, {de
  Martino}, {de Souza}, {Debosscher}, {del Pozo}, {Delbo}, {Delgado},
  {Delgado}, {di Marco}, {Di Matteo}, {Diakite}, {Distefano}, {Dolding}, {Dos
  Anjos}, {Drazinos}, {Dur{\'a}n}, {Dzigan}, {Ecale}, {Edvardsson}, {Enke},
  {Erdmann}, {Escolar}, {Espina}, {Evans}, {Eynard Bontemps}, {Fabre},
  {Fabrizio}, {Faigler}, {Falc{\~a}o}, {Farr{\`a}s Casas}, {Faye}, {Federici},
  {Fedorets}, {Fern{\'a}ndez-Hern{\'a}ndez}, {Fernique}, {Fienga}, {Figueras},
  {Filippi}, {Findeisen}, {Fonti}, {Fouesneau}, {Fraile}, {Fraser}, {Fuchs},
  {Furnell}, {Gai}, {Galleti}, {Galluccio}, {Garabato}, {Garc{\'\i}a-Sedano},
  {Gar{\'e}}, {Garofalo}, {Garralda}, {Gavras}, {Gerssen}, {Geyer}, {Gilmore},
  {Girona}, {Giuffrida}, {Gomes}, {Gonz{\'a}lez-Marcos},
  {Gonz{\'a}lez-N{\'u}{\~n}ez}, {Gonz{\'a}lez-Vidal}, {Granvik}, {Guerrier},
  {Guillout}, {Guiraud}, {G{\'u}rpide}, {Guti{\'e}rrez-S{\'a}nchez}, {Guy},
  {Haigron}, {Hatzidimitriou}, {Haywood}, {Heiter}, {Helmi}, {Hobbs},
  {Hofmann}, {Holl}, {Holland}, {Hunt}, {Hypki}, {Icardi}, {Irwin}, {Jevardat
  de Fombelle}, {Jofr{\'e}}, {Jonker}, {Jorissen}, {Julbe}, {Karampelas},
  {Kochoska}, {Kohley}, {Kolenberg}, {Kontizas}, {Koposov}, {Kordopatis},
  {Koubsky}, {Kowalczyk}, {Krone-Martins}, {Kudryashova}, {Kull}, {Bachchan},
  {Lacoste-Seris}, {Lanza}, {Lavigne}, {Le Poncin-Lafitte}, {Lebreton},
  {Lebzelter}, {Leccia}, {Leclerc}, {Lecoeur-Taibi}, {Lemaitre}, {Lenhardt},
  {Leroux}, {Liao}, {Licata}, {Lindstr{\o}m}, {Lister}, {Livanou}, {Lobel},
  {L{\"o}ffler}, {L{\'o}pez}, {Lopez-Lozano}, {Lorenz}, {Loureiro},
  {MacDonald}, {Magalh{\~a}es Fernandes}, {Managau}, {Mann}, {Mantelet},
  {Marchal}, {Marchant}, {Marconi}, {Marie}, {Marinoni}, {Marrese},
  {Marschalk{\'o}}, {Marshall}, {Mart{\'\i}n-Fleitas}, {Martino}, {Mary},
  {Matijevi{\v{c}}}, {Mazeh}, {McMillan}, {Messina}, {Mestre}, {Michalik},
  {Millar}, {Miranda}, {Molina}, {Molinaro}, {Molinaro}, {Moln{\'a}r},
  {Moniez}, {Montegriffo}, {Monteiro}, {Mor}, {Mora}, {Morbidelli}, {Morel},
  {Morgenthaler}, {Morley}, {Morris}, {Mulone}, {Muraveva}, {Musella},
  {Narbonne}, {Nelemans}, {Nicastro}, {Noval}, {Ord{\'e}novic},
  {Ordieres-Mer{\'e}}, {Osborne}, {Pagani}, {Pagano}, {Pailler}, {Palacin},
  {Palaversa}, {Parsons}, {Paulsen}, {Pecoraro}, {Pedrosa}, {Pentik{\"a}inen},
  {Pereira}, {Pichon}, {Piersimoni}, {Pineau}, {Plachy}, {Plum}, {Poujoulet},
  {Pr{\v{s}}a}, {Pulone}, {Ragaini}, {Rago}, {Rambaux}, {Ramos-Lerate},
  {Ranalli}, {Rauw}, {Read}, {Regibo}, {Renk}, {Reyl{\'e}}, {Ribeiro},
  {Rimoldini}, {Ripepi}, {Riva}, {Rixon}, {Roelens}, {Romero-G{\'o}mez},
  {Rowell}, {Royer}, {Rudolph}, {Ruiz-Dern}, {Sadowski}, {Sagrist{\`a}
  Sell{\'e}s}, {Sahlmann}, {Salgado}, {Salguero}, {Sarasso}, {Savietto},
  {Schnorhk}, {Schultheis}, {Sciacca}, {Segol}, {Segovia}, {Segransan},
  {Serpell}, {Shih}, {Smareglia}, {Smart}, {Smith}, {Solano}, {Solitro},
  {Sordo}, {Soria Nieto}, {Souchay}, {Spagna}, {Spoto}, {Stampa}, {Steele},
  {Steidelm{\"u}ller}, {Stephenson}, {Stoev}, {Suess}, {S{\"u}veges}, {Surdej},
  {Szabados}, {Szegedi-Elek}, {Tapiador}, {Taris}, {Tauran}, {Taylor},
  {Teixeira}, {Terrett}, {Tingley}, {Trager}, {Turon}, {Ulla}, {Utrilla},
  {Valentini}, {van Elteren}, {Van Hemelryck}, {van Leeuwen}, {Varadi},
  {Vecchiato}, {Veljanoski}, {Via}, {Vicente}, {Vogt}, {Voss}, {Votruba},
  {Voutsinas}, {Walmsley}, {Weiler}, {Weingrill}, {Werner}, {Wevers},
  {Whitehead}, {Wyrzykowski}, {Yoldas}, {{\v{Z}}erjal}, {Zucker}, {Zurbach},
  {Zwitter}, {Alecu}, {Allen}, {Allende Prieto}, {Amorim},
  {Anglada-Escud{\'e}}, {Arsenijevic}, {Azaz}, {Balm}, {Beck}, {Bernstein},
  {Bigot}, {Bijaoui}, {Blasco}, {Bonfigli}, {Bono}, {Boudreault}, {Bressan},
  {Brown}, {Brunet}, {Bunclark}, {Buonanno}, {Butkevich}, {Carret}, {Carrion},
  {Chemin}, {Ch{\'e}reau}, {Corcione}, {Darmigny}, {de Boer}, {de Teodoro}, {de
  Zeeuw}, {Delle Luche}, {Domingues}, {Dubath}, {Fodor}, {Fr{\'e}zouls},
  {Fries}, {Fustes}, {Fyfe}, {Gallardo}, {Gallegos}, {Gardiol}, {Gebran},
  {Gomboc}, {G{\'o}mez}, {Grux}, {Gueguen}, {Heyrovsky}, {Hoar}, {Iannicola},
  {Isasi Parache}, {Janotto}, {Joliet}, {Jonckheere}, {Keil}, {Kim},
  {Klagyivik}, {Klar}, {Knude}, {Kochukhov}, {Kolka}, {Kos}, {Kutka}, {Lainey},
  {LeBouquin}, {Liu}, {Loreggia}, {Makarov}, {Marseille}, {Martayan},
  {Martinez-Rubi}, {Massart}, {Meynadier}, {Mignot}, {Munari}, {Nguyen},
  {Nordlander}, {Ocvirk}, {O'Flaherty}, {Olias Sanz}, {Ortiz}, {Osorio},
  {Oszkiewicz}, {Ouzounis}, {Palmer}, {Park}, {Pasquato}, {Peltzer}, {Peralta},
  {P{\'e}turaud}, {Pieniluoma}, {Pigozzi}, {Poels}, {Prat}, {Prod'homme},
  {Raison}, {Rebordao}, {Risquez}, {Rocca-Volmerange}, {Rosen}, {Ruiz-Fuertes},
  {Russo}, {Sembay}, {Serraller Vizcaino}, {Short}, {Siebert}, {Silva},
  {Sinachopoulos}, {Slezak}, {Soffel}, {Sosnowska}, {Strai{\v{z}}ys}, {ter
  Linden}, {Terrell}, {Theil}, {Tiede}, {Troisi}, {Tsalmantza}, {Tur},
  {Vaccari}, {Vachier}, {Valles}, {Van Hamme}, {Veltz}, {Virtanen}, {Wallut},
  {Wichmann}, {Wilkinson}, {Ziaeepour}, \& {Zschocke}}]{Gaia2016}
{Gaia Collaboration}, {Prusti}, T., {de Bruijne}, J.~H.~J., {et~al.} 2016,
  \bibinfo{title}{{The Gaia mission},} \aap, 595, A1,
  \dodoi{10.1051/0004-6361/201629272}

\bibitem[{ {Gaia Collaboration} {et~al.}(2018){Gaia Collaboration}, {Brown},
  {Vallenari}, {Prusti}, {de Bruijne}, {Babusiaux}, {Bailer-Jones}, {Biermann},
  {Evans}, {Eyer}, {Jansen}, {Jordi}, {Klioner}, {Lammers}, {Lindegren},
  {Luri}, {Mignard}, {Panem}, {Pourbaix}, {Randich}, {Sartoretti}, {Siddiqui},
  {Soubiran}, {van Leeuwen}, {Walton}, {Arenou}, {Bastian}, {Cropper},
  {Drimmel}, {Katz}, {Lattanzi}, {Bakker}, {Cacciari}, {Casta{\~n}eda},
  {Chaoul}, {Cheek}, {De Angeli}, {Fabricius}, {Guerra}, {Holl}, {Masana},
  {Messineo}, {Mowlavi}, {Nienartowicz}, {Panuzzo}, {Portell}, {Riello},
  {Seabroke}, {Tanga}, {Th{\'e}venin}, {Gracia-Abril}, {Comoretto},
  {Garcia-Reinaldos}, {Teyssier}, {Altmann}, {Andrae}, {Audard},
  {Bellas-Velidis}, {Benson}, {Berthier}, {Blomme}, {Burgess}, {Busso},
  {Carry}, {Cellino}, {Clementini}, {Clotet}, {Creevey}, {Davidson}, {De
  Ridder}, {Delchambre}, {Dell'Oro}, {Ducourant},
  {Fern{\'a}ndez-Hern{\'a}ndez}, {Fouesneau}, {Fr{\'e}mat}, {Galluccio},
  {Garc{\'\i}a-Torres}, {Gonz{\'a}lez-N{\'u}{\~n}ez}, {Gonz{\'a}lez-Vidal},
  {Gosset}, {Guy}, {Halbwachs}, {Hambly}, {Harrison}, {Hern{\'a}ndez},
  {Hestroffer}, {Hodgkin}, {Hutton}, {Jasniewicz}, {Jean-Antoine-Piccolo},
  {Jordan}, {Korn}, {Krone-Martins}, {Lanzafame}, {Lebzelter}, {L{\"o}ffler},
  {Manteiga}, {Marrese}, {Mart{\'\i}n-Fleitas}, {Moitinho}, {Mora}, {Muinonen},
  {Osinde}, {Pancino}, {Pauwels}, {Petit}, {Recio-Blanco}, {Richards},
  {Rimoldini}, {Robin}, {Sarro}, {Siopis}, {Smith}, {Sozzetti}, {S{\"u}veges},
  {Torra}, {van Reeven}, {Abbas}, {Abreu Aramburu}, {Accart}, {Aerts},
  {Altavilla}, {{\'A}lvarez}, {Alvarez}, {Alves}, {Anderson}, {Andrei},
  {Anglada Varela}, {Antiche}, {Antoja}, {Arcay}, {Astraatmadja}, {Bach},
  {Baker}, {Balaguer-N{\'u}{\~n}ez}, {Balm}, {Barache}, {Barata}, {Barbato},
  {Barblan}, {Barklem}, {Barrado}, {Barros}, {Barstow}, {Bartholom{\'e}
  Mu{\~n}oz}, {Bassilana}, {Becciani}, {Bellazzini}, {Berihuete}, {Bertone},
  {Bianchi}, {Bienaym{\'e}}, {Blanco-Cuaresma}, {Boch}, {Boeche}, {Bombrun},
  {Borrachero}, {Bossini}, {Bouquillon}, {Bourda}, {Bragaglia}, {Bramante},
  {Breddels}, {Bressan}, {Brouillet}, {Br{\"u}semeister}, {Brugaletta},
  {Bucciarelli}, {Burlacu}, {Busonero}, {Butkevich}, {Buzzi}, {Caffau},
  {Cancelliere}, {Cannizzaro}, {Cantat-Gaudin}, {Carballo}, {Carlucci},
  {Carrasco}, {Casamiquela}, {Castellani}, {Castro-Ginard}, {Charlot},
  {Chemin}, {Chiavassa}, {Cocozza}, {Costigan}, {Cowell}, {Crifo}, {Crosta},
  {Crowley}, {Cuypers}, {Dafonte}, {Damerdji}, {Dapergolas}, {David}, {David},
  {de Laverny}, {De Luise}, {De March}, {de Martino}, {de Souza}, {de Torres},
  {Debosscher}, {del Pozo}, {Delbo}, {Delgado}, {Delgado}, {Di Matteo},
  {Diakite}, {Diener}, {Distefano}, {Dolding}, {Drazinos}, {Dur{\'a}n},
  {Edvardsson}, {Enke}, {Eriksson}, {Esquej}, {Eynard Bontemps}, {Fabre},
  {Fabrizio}, {Faigler}, {Falc{\~a}o}, {Farr{\`a}s Casas}, {Federici},
  {Fedorets}, {Fernique}, {Figueras}, {Filippi}, {Findeisen}, {Fonti},
  {Fraile}, {Fraser}, {Fr{\'e}zouls}, {Gai}, {Galleti}, {Garabato},
  {Garc{\'\i}a-Sedano}, {Garofalo}, {Garralda}, {Gavel}, {Gavras}, {Gerssen},
  {Geyer}, {Giacobbe}, {Gilmore}, {Girona}, {Giuffrida}, {Glass}, {Gomes},
  {Granvik}, {Gueguen}, {Guerrier}, {Guiraud}, {Guti{\'e}rrez-S{\'a}nchez},
  {Haigron}, {Hatzidimitriou}, {Hauser}, {Haywood}, {Heiter}, {Helmi}, {Heu},
  {Hilger}, {Hobbs}, {Hofmann}, {Holland}, {Huckle}, {Hypki}, {Icardi},
  {Jan{\ss}en}, {Jevardat de Fombelle}, {Jonker}, {Juh{\'a}sz}, {Julbe},
  {Karampelas}, {Kewley}, {Klar}, {Kochoska}, {Kohley}, {Kolenberg},
  {Kontizas}, {Kontizas}, {Koposov}, {Kordopatis}, {Kostrzewa-Rutkowska},
  {Koubsky}, {Lambert}, {Lanza}, {Lasne}, {Lavigne}, {Le Fustec}, {Le
  Poncin-Lafitte}, {Lebreton}, {Leccia}, {Leclerc}, {Lecoeur-Taibi},
  {Lenhardt}, {Leroux}, {Liao}, {Licata}, {Lindstr{\o}m}, {Lister}, {Livanou},
  {Lobel}, {L{\'o}pez}, {Managau}, {Mann}, {Mantelet}, {Marchal}, {Marchant},
  {Marconi}, {Marinoni}, {Marschalk{\'o}}, {Marshall}, {Martino}, {Marton},
  {Mary}, {Massari}, {Matijevi{\v{c}}}, {Mazeh}, {McMillan}, {Messina},
  {Michalik}, {Millar}, {Molina}, {Molinaro}, {Moln{\'a}r}, {Montegriffo},
  {Mor}, {Morbidelli}, {Morel}, {Morris}, {Mulone}, {Muraveva}, {Musella},
  {Nelemans}, {Nicastro}, {Noval}, {O'Mullane}, {Ord{\'e}novic},
  {Ord{\'o}{\~n}ez-Blanco}, {Osborne}, {Pagani}, {Pagano}, {Pailler},
  {Palacin}, {Palaversa}, {Panahi}, {Pawlak}, {Piersimoni}, {Pineau}, {Plachy},
  {Plum}, {Poggio}, {Poujoulet}, {Pr{\v{s}}a}, {Pulone}, {Racero}, {Ragaini},
  {Rambaux}, {Ramos-Lerate}, {Regibo}, {Reyl{\'e}}, {Riclet}, {Ripepi}, {Riva},
  {Rivard}, {Rixon}, {Roegiers}, {Roelens}, {Romero-G{\'o}mez}, {Rowell},
  {Royer}, {Ruiz-Dern}, {Sadowski}, {Sagrist{\`a} Sell{\'e}s}, {Sahlmann},
  {Salgado}, {Salguero}, {Sanna}, {Santana-Ros}, {Sarasso}, {Savietto},
  {Schultheis}, {Sciacca}, {Segol}, {Segovia}, {S{\'e}gransan}, {Shih},
  {Siltala}, {Silva}, {Smart}, {Smith}, {Solano}, {Solitro}, {Sordo}, {Soria
  Nieto}, {Souchay}, {Spagna}, {Spoto}, {Stampa}, {Steele},
  {Steidelm{\"u}ller}, {Stephenson}, {Stoev}, {Suess}, {Surdej}, {Szabados},
  {Szegedi-Elek}, {Tapiador}, {Taris}, {Tauran}, {Taylor}, {Teixeira},
  {Terrett}, {Teyssandier}, {Thuillot}, {Titarenko}, {Torra Clotet}, {Turon},
  {Ulla}, {Utrilla}, {Uzzi}, {Vaillant}, {Valentini}, {Valette}, {van Elteren},
  {Van Hemelryck}, {van Leeuwen}, {Vaschetto}, {Vecchiato}, {Veljanoski},
  {Viala}, {Vicente}, {Vogt}, {von Essen}, {Voss}, {Votruba}, {Voutsinas},
  {Walmsley}, {Weiler}, {Wertz}, {Wevers}, {Wyrzykowski}, {Yoldas},
  {{\v{Z}}erjal}, {Ziaeepour}, {Zorec}, {Zschocke}, {Zucker}, {Zurbach}, \&
  {Zwitter}}]{Gaia2018}
{Gaia Collaboration}, {Brown}, A.~G.~A., {Vallenari}, A., {et~al.} 2018,
  \bibinfo{title}{{Gaia Data Release 2. Summary of the contents and survey
  properties},} \aap, 616, A1, \dodoi{10.1051/0004-6361/201833051}

\bibitem[{ {Gaia Collaboration} {et~al.}(2023){Gaia Collaboration},
  {Vallenari}, {Brown}, {Prusti}, {de Bruijne}, {Arenou}, {Babusiaux},
  {Biermann}, {Creevey}, {Ducourant}, {Evans}, {Eyer}, {Guerra}, {Hutton},
  {Jordi}, {Klioner}, {Lammers}, {Lindegren}, {Luri}, {Mignard}, {Panem},
  {Pourbaix}, {Randich}, {Sartoretti}, {Soubiran}, {Tanga}, {Walton},
  {Bailer-Jones}, {Bastian}, {Drimmel}, {Jansen}, {Katz}, {Lattanzi}, {van
  Leeuwen}, {Bakker}, {Cacciari}, {Casta{\~n}eda}, {De Angeli}, {Fabricius},
  {Fouesneau}, {Fr{\'e}mat}, {Galluccio}, {Guerrier}, {Heiter}, {Masana},
  {Messineo}, {Mowlavi}, {Nicolas}, {Nienartowicz}, {Pailler}, {Panuzzo},
  {Riclet}, {Roux}, {Seabroke}, {Sordo}, {Th{\'e}venin}, {Gracia-Abril},
  {Portell}, {Teyssier}, {Altmann}, {Andrae}, {Audard}, {Bellas-Velidis},
  {Benson}, {Berthier}, {Blomme}, {Burgess}, {Busonero}, {Busso},
  {C{\'a}novas}, {Carry}, {Cellino}, {Cheek}, {Clementini}, {Damerdji},
  {Davidson}, {de Teodoro}, {Nu{\~n}ez Campos}, {Delchambre}, {Dell'Oro},
  {Esquej}, {Fern{\'a}ndez-Hern{\'a}ndez}, {Fraile}, {Garabato},
  {Garc{\'\i}a-Lario}, {Gosset}, {Haigron}, {Halbwachs}, {Hambly}, {Harrison},
  {Hern{\'a}ndez}, {Hestroffer}, {Hodgkin}, {Holl}, {Jan{\ss}en}, {Jevardat de
  Fombelle}, {Jordan}, {Krone-Martins}, {Lanzafame}, {L{\"o}ffler}, {Marchal},
  {Marrese}, {Moitinho}, {Muinonen}, {Osborne}, {Pancino}, {Pauwels},
  {Recio-Blanco}, {Reyl{\'e}}, {Riello}, {Rimoldini}, {Roegiers}, {Rybizki},
  {Sarro}, {Siopis}, {Smith}, {Sozzetti}, {Utrilla}, {van Leeuwen}, {Abbas},
  {{\'A}brah{\'a}m}, {Abreu Aramburu}, {Aerts}, {Aguado}, {Ajaj},
  {Aldea-Montero}, {Altavilla}, {{\'A}lvarez}, {Alves}, {Anders}, {Anderson},
  {Anglada Varela}, {Antoja}, {Baines}, {Baker}, {Balaguer-N{\'u}{\~n}ez},
  {Balbinot}, {Balog}, {Barache}, {Barbato}, {Barros}, {Barstow},
  {Bartolom{\'e}}, {Bassilana}, {Bauchet}, {Becciani}, {Bellazzini},
  {Berihuete}, {Bernet}, {Bertone}, {Bianchi}, {Binnenfeld}, {Blanco-Cuaresma},
  {Blazere}, {Boch}, {Bombrun}, {Bossini}, {Bouquillon}, {Bragaglia},
  {Bramante}, {Breedt}, {Bressan}, {Brouillet}, {Brugaletta}, {Bucciarelli},
  {Burlacu}, {Butkevich}, {Buzzi}, {Caffau}, {Cancelliere}, {Cantat-Gaudin},
  {Carballo}, {Carlucci}, {Carnerero}, {Carrasco}, {Casamiquela}, {Castellani},
  {Castro-Ginard}, {Chaoul}, {Charlot}, {Chemin}, {Chiaramida}, {Chiavassa},
  {Chornay}, {Comoretto}, {Contursi}, {Cooper}, {Cornez}, {Cowell}, {Crifo},
  {Cropper}, {Crosta}, {Crowley}, {Dafonte}, {Dapergolas}, {David}, {David},
  {de Laverny}, {De Luise}, {De March}, {De Ridder}, {de Souza}, {de Torres},
  {del Peloso}, {del Pozo}, {Delbo}, {Delgado}, {Delisle}, {Demouchy},
  {Dharmawardena}, {Di Matteo}, {Diakite}, {Diener}, {Distefano}, {Dolding},
  {Edvardsson}, {Enke}, {Fabre}, {Fabrizio}, {Faigler}, {Fedorets}, {Fernique},
  {Fienga}, {Figueras}, {Fournier}, {Fouron}, {Fragkoudi}, {Gai},
  {Garcia-Gutierrez}, {Garcia-Reinaldos}, {Garc{\'\i}a-Torres}, {Garofalo},
  {Gavel}, {Gavras}, {Gerlach}, {Geyer}, {Giacobbe}, {Gilmore}, {Girona},
  {Giuffrida}, {Gomel}, {Gomez}, {Gonz{\'a}lez-N{\'u}{\~n}ez},
  {Gonz{\'a}lez-Santamar{\'\i}a}, {Gonz{\'a}lez-Vidal}, {Granvik}, {Guillout},
  {Guiraud}, {Guti{\'e}rrez-S{\'a}nchez}, {Guy}, {Hatzidimitriou}, {Hauser},
  {Haywood}, {Helmer}, {Helmi}, {Sarmiento}, {Hidalgo}, {Hilger},
  {H{\l}adczuk}, {Hobbs}, {Holland}, {Huckle}, {Jardine}, {Jasniewicz},
  {Jean-Antoine Piccolo}, {Jim{\'e}nez-Arranz}, {Jorissen}, {Juaristi
  Campillo}, {Julbe}, {Karbevska}, {Kervella}, {Khanna}, {Kontizas},
  {Kordopatis}, {Korn}, {K{\'o}sp{\'a}l}, {Kostrzewa-Rutkowska},
  {Kruszy{\'n}ska}, {Kun}, {Laizeau}, {Lambert}, {Lanza}, {Lasne}, {Le
  Campion}, {Lebreton}, {Lebzelter}, {Leccia}, {Leclerc}, {Lecoeur-Taibi},
  {Liao}, {Licata}, {Lindstr{\o}m}, {Lister}, {Livanou}, {Lobel}, {Lorca},
  {Loup}, {Madrero Pardo}, {Magdaleno Romeo}, {Managau}, {Mann}, {Manteiga},
  {Marchant}, {Marconi}, {Marcos}, {Marcos Santos}, {Mar{\'\i}n Pina},
  {Marinoni}, {Marocco}, {Marshall}, {Martin Polo}, {Mart{\'\i}n-Fleitas},
  {Marton}, {Mary}, {Masip}, {Massari}, {Mastrobuono-Battisti}, {Mazeh},
  {McMillan}, {Messina}, {Michalik}, {Millar}, {Mints}, {Molina}, {Molinaro},
  {Moln{\'a}r}, {Monari}, {Mongui{\'o}}, {Montegriffo}, {Montero}, {Mor},
  {Mora}, {Morbidelli}, {Morel}, {Morris}, {Muraveva}, {Murphy}, {Musella},
  {Nagy}, {Noval}, {Oca{\~n}a}, {Ogden}, {Ordenovic}, {Osinde}, {Pagani},
  {Pagano}, {Palaversa}, {Palicio}, {Pallas-Quintela}, {Panahi},
  {Payne-Wardenaar}, {Pe{\~n}alosa Esteller}, {Penttil{\"a}}, {Pichon},
  {Piersimoni}, {Pineau}, {Plachy}, {Plum}, {Poggio}, {Pr{\v{s}}a}, {Pulone},
  {Racero}, {Ragaini}, {Rainer}, {Raiteri}, {Rambaux}, {Ramos}, {Ramos-Lerate},
  {Re Fiorentin}, {Regibo}, {Richards}, {Rios Diaz}, {Ripepi}, {Riva}, {Rix},
  {Rixon}, {Robichon}, {Robin}, {Robin}, {Roelens}, {Rogues}, {Rohrbasser},
  {Romero-G{\'o}mez}, {Rowell}, {Royer}, {Ruz Mieres}, {Rybicki}, {Sadowski},
  {S{\'a}ez N{\'u}{\~n}ez}, {Sagrist{\`a} Sell{\'e}s}, {Sahlmann}, {Salguero},
  {Samaras}, {Sanchez Gimenez}, {Sanna}, {Santove{\~n}a}, {Sarasso},
  {Schultheis}, {Sciacca}, {Segol}, {Segovia}, {S{\'e}gransan}, {Semeux},
  {Shahaf}, {Siddiqui}, {Siebert}, {Siltala}, {Silvelo}, {Slezak}, {Slezak},
  {Smart}, {Snaith}, {Solano}, {Solitro}, {Souami}, {Souchay}, {Spagna},
  {Spina}, {Spoto}, {Steele}, {Steidelm{\"u}ller}, {Stephenson}, {S{\"u}veges},
  {Surdej}, {Szabados}, {Szegedi-Elek}, {Taris}, {Taylor}, {Teixeira},
  {Tolomei}, {Tonello}, {Torra}, {Torra}, {Torralba Elipe}, {Trabucchi},
  {Tsounis}, {Turon}, {Ulla}, {Unger}, {Vaillant}, {van Dillen}, {van Reeven},
  {Vanel}, {Vecchiato}, {Viala}, {Vicente}, {Voutsinas}, {Weiler}, {Wevers},
  {Wyrzykowski}, {Yoldas}, {Yvard}, {Zhao}, {Zorec}, {Zucker}, \&
  {Zwitter}}]{Gaia2023}
{Gaia Collaboration}, {Vallenari}, A., {Brown}, A.~G.~A., {et~al.} 2023,
  \bibinfo{title}{{Gaia Data Release 3. Summary of the content and survey
  properties},} \aap, 674, A1, \dodoi{10.1051/0004-6361/202243940}

\bibitem[{E. {Gaidos} {et~al.}(2016){Gaidos}, {Mann}, {Kraus}, \&
  {Ireland}}]{Gaidos2016}
{Gaidos}, E., {Mann}, A.~W., {Kraus}, A.~L., \& {Ireland}, M. 2016,
  \bibinfo{title}{{They are small worlds after all: revised properties of
  Kepler M dwarf stars and their planets},} \mnras, 457, 2877,
  \dodoi{10.1093/mnras/stw097}

\bibitem[{G.~J. {Gilbert} {et~al.}(2025){Gilbert}, {Petigura}, \&
  {Entrican}}]{Gilbert2025}
{Gilbert}, G.~J., {Petigura}, E.~A., \& {Entrican}, P.~M. 2025,
  \bibinfo{title}{{Planets larger than Neptune have elevated eccentricities},}
  Proceedings of the National Academy of Science, 122, e2405295122,
  \dodoi{10.1073/pnas.2405295122}

\bibitem[{C.~R. Harris {et~al.}(2020)Harris, Millman, van~der Walt, Gommers,
  Virtanen, Cournapeau, Wieser, Taylor, Berg, Smith, Kern, Picus, Hoyer, van
  Kerkwijk, Brett, Haldane, Fern{\'a}ndez~del R{\'\i}o, Wiebe, Peterson,
  G{\'e}rard-Marchant, Sheppard, Reddy, Weckesser, Abbasi, Gohlke, \&
  Oliphant}]{Harris2020}
Harris, C.~R., Millman, K.~J., van~der Walt, S.~J., {et~al.} 2020,
  \bibinfo{title}{Array programming with {NumPy},} Nature, 585, 357,
  \dodoi{10.1038/s41586-020-2649-2}

\bibitem[{R.~J. {Harris} {et~al.}(2012){Harris}, {Andrews}, {Wilner}, \&
  {Kraus}}]{Harris2012}
{Harris}, R.~J., {Andrews}, S.~M., {Wilner}, D.~J., \& {Kraus}, A.~L. 2012,
  \bibinfo{title}{{A Resolved Census of Millimeter Emission from Taurus
  Multiple Star Systems},} \apj, 751, 115, \dodoi{10.1088/0004-637X/751/2/115}

\bibitem[{L.~A. {Hirsch} {et~al.}(2021){Hirsch}, {Rosenthal}, {Fulton},
  {Howard}, {Ciardi}, {Marcy}, {Nielsen}, {Petigura}, {de Rosa}, {Isaacson},
  {Weiss}, {Sinukoff}, \& {Macintosh}}]{Hirsch2021}
{Hirsch}, L.~A., {Rosenthal}, L., {Fulton}, B.~J., {et~al.} 2021,
  \bibinfo{title}{{Understanding the Impacts of Stellar Companions on Planet
  Formation and Evolution: A Survey of Stellar and Planetary Companions within
  25 pc},} \aj, 161, 134, \dodoi{10.3847/1538-3881/abd639}

\bibitem[{C.~S.~K. {Ho} {et~al.}(2024){Ho}, {Rogers}, {Van Eylen}, {Owen}, \&
  {Schlichting}}]{Ho2024}
{Ho}, C. S.~K., {Rogers}, J.~G., {Van Eylen}, V., {Owen}, J.~E., \&
  {Schlichting}, H.~E. 2024, \bibinfo{title}{{Shallower radius valley around
  low-mass hosts: evidence for icy planets, collisions, or high-energy
  radiation scatter},} \mnras, 531, 3698, \dodoi{10.1093/mnras/stae1376}

\bibitem[{C.~S.~K. {Ho} \& V. {Van Eylen}(2023){Ho} \& {Van Eylen}}]{Ho2023}
{Ho}, C. S.~K., \& {Van Eylen}, V. 2023, \bibinfo{title}{{A deep radius valley
  revealed by Kepler short cadence observations},} \mnras, 519, 4056,
  \dodoi{10.1093/mnras/stac3802}

\bibitem[{A.~W. {Howard} {et~al.}(2012){Howard}, {Marcy}, {Bryson}, {Jenkins},
  {Rowe}, {Batalha}, {Borucki}, {Koch}, {Dunham}, {Gautier}, {Van Cleve},
  {Cochran}, {Latham}, {Lissauer}, {Torres}, {Brown}, {Gilliland}, {Buchhave},
  {Caldwell}, {Christensen-Dalsgaard}, {Ciardi}, {Fressin}, {Haas}, {Howell},
  {Kjeldsen}, {Seager}, {Rogers}, {Sasselov}, {Steffen}, {Basri},
  {Charbonneau}, {Christiansen}, {Clarke}, {Dupree}, {Fabrycky}, {Fischer},
  {Ford}, {Fortney}, {Tarter}, {Girouard}, {Holman}, {Johnson}, {Klaus},
  {Machalek}, {Moorhead}, {Morehead}, {Ragozzine}, {Tenenbaum}, {Twicken},
  {Quinn}, {Isaacson}, {Shporer}, {Lucas}, {Walkowicz}, {Welsh}, {Boss},
  {Devore}, {Gould}, {Smith}, {Morris}, {Prsa}, {Morton}, {Still}, {Thompson},
  {Mullally}, {Endl}, \& {MacQueen}}]{Howard2012}
{Howard}, A.~W., {Marcy}, G.~W., {Bryson}, S.~T., {et~al.} 2012,
  \bibinfo{title}{{Planet Occurrence within 0.25 AU of Solar-type Stars from
  Kepler},} \apjs, 201, 15, \dodoi{10.1088/0067-0049/201/2/15}

\bibitem[{S.~B. {Howell} {et~al.}(2011){Howell}, {Everett}, {Sherry}, {Horch},
  \& {Ciardi}}]{Howell2011}
{Howell}, S.~B., {Everett}, M.~E., {Sherry}, W., {Horch}, E., \& {Ciardi},
  D.~R. 2011, \bibinfo{title}{{Speckle Camera Observations for the NASA Kepler
  Mission Follow-up Program},} \aj, 142, 19, \dodoi{10.1088/0004-6256/142/1/19}

\bibitem[{J.~D. {Hunter}(2007){Hunter}}]{Hunter2007}
{Hunter}, J.~D. 2007, \bibinfo{title}{{Matplotlib: A 2D Graphics Environment},}
  Computing in Science and Engineering, 9, 90, \dodoi{10.1109/MCSE.2007.55}

\bibitem[{H. {Jang-Condell}(2015){Jang-Condell}}]{JangCondell2015}
{Jang-Condell}, H. 2015, \bibinfo{title}{{On the Likelihood of Planet Formation
  in Close Binaries},} \apj, 799, 147, \dodoi{10.1088/0004-637X/799/2/147}

\bibitem[{ {Kepler Mission}(2019){Kepler Mission}}]{Planet_catalog_DOI}
{Kepler Mission}. 2019, \bibinfo{title}{Kepler Objects of Interest DR 25
  Table,} IPAC, \dodoi{10.26133/NEA5}

\bibitem[{ {Kepler Project}(2020){Kepler Project}}]{Kepler_completeness_DOI}
{Kepler Project}. 2020, \bibinfo{title}{Kepler Mission Completeness and
  Reliability Products,} IPAC, \dodoi{10.26133/NEA14}

\bibitem[{A.~L. {Kraus} {et~al.}(2012){Kraus}, {Ireland}, {Hillenbrand}, \&
  {Martinache}}]{Kraus2012}
{Kraus}, A.~L., {Ireland}, M.~J., {Hillenbrand}, L.~A., \& {Martinache}, F.
  2012, \bibinfo{title}{{The Role of Multiplicity in Disk Evolution and Planet
  Formation},} \apj, 745, 19, \dodoi{10.1088/0004-637X/745/1/19}

\bibitem[{A.~L. {Kraus} {et~al.}(2016){Kraus}, {Ireland}, {Huber}, {Mann}, \&
  {Dupuy}}]{Kraus2016}
{Kraus}, A.~L., {Ireland}, M.~J., {Huber}, D., {Mann}, A.~W., \& {Dupuy}, T.~J.
  2016, \bibinfo{title}{{The Impact of Stellar Multiplicity on Planetary
  Systems. I. The Ruinous Influence of Close Binary Companions},} \aj, 152, 8,
  \dodoi{10.3847/0004-6256/152/1/8}

\bibitem[{M. {Kunimoto} \& J.~M. {Matthews}(2020){Kunimoto} \&
  {Matthews}}]{Kunimoto2020}
{Kunimoto}, M., \& {Matthews}, J.~M. 2020, \bibinfo{title}{{Searching the
  Entirety of Kepler Data. II. Occurrence Rate Estimates for FGK Stars},} \aj,
  159, 248, \dodoi{10.3847/1538-3881/ab88b0}

\bibitem[{K.~V. {Lester} {et~al.}(2021){Lester}, {Matson}, {Howell}, {Furlan},
  {Gnilka}, {Scott}, {Ciardi}, {Everett}, {Hartman}, \& {Hirsch}}]{Lester2021}
{Lester}, K.~V., {Matson}, R.~A., {Howell}, S.~B., {et~al.} 2021,
  \bibinfo{title}{{Speckle Observations of TESS Exoplanet Host Stars. II.
  Stellar Companions at 1-1000 au and Implications for Small Planet
  Detection},} \aj, 162, 75, \dodoi{10.3847/1538-3881/ac0d06}

\bibitem[{J. {Lillo-Box} {et~al.}(2012){Lillo-Box}, {Barrado}, \&
  {Bouy}}]{Lillo-Box2012}
{Lillo-Box}, J., {Barrado}, D., \& {Bouy}, H. 2012,
  \bibinfo{title}{{Multiplicity in transiting planet-host stars. A lucky
  imaging study of Kepler candidates},} \aap, 546, A10,
  \dodoi{10.1051/0004-6361/201219631}

\bibitem[{J. {Lillo-Box} {et~al.}(2014){Lillo-Box}, {Barrado}, \&
  {Bouy}}]{Lillo-Box2014}
{Lillo-Box}, J., {Barrado}, D., \& {Bouy}, H. 2014,
  \bibinfo{title}{{High-resolution imaging of Kepler planet host candidates. A
  comprehensive comparison of different techniques},} \aap, 566, A103,
  \dodoi{10.1051/0004-6361/201423497}

\bibitem[{L. {Lindegren} {et~al.}(2021){Lindegren}, {Klioner}, {Hern{\'a}ndez},
  {Bombrun}, {Ramos-Lerate}, {Steidelm{\"u}ller}, {Bastian}, {Biermann}, {de
  Torres}, {Gerlach}, {Geyer}, {Hilger}, {Hobbs}, {Lammers}, {McMillan},
  {Stephenson}, {Casta{\~n}eda}, {Davidson}, {Fabricius}, {Gracia-Abril},
  {Portell}, {Rowell}, {Teyssier}, {Torra}, {Bartolom{\'e}}, {Clotet},
  {Garralda}, {Gonz{\'a}lez-Vidal}, {Torra}, {Abbas}, {Altmann}, {Anglada
  Varela}, {Balaguer-N{\'u}{\~n}ez}, {Balog}, {Barache}, {Becciani}, {Bernet},
  {Bertone}, {Bianchi}, {Bouquillon}, {Brown}, {Bucciarelli}, {Busonero},
  {Butkevich}, {Buzzi}, {Cancelliere}, {Carlucci}, {Charlot}, {Cioni},
  {Crosta}, {Crowley}, {del Peloso}, {del Pozo}, {Drimmel}, {Esquej}, {Fienga},
  {Fraile}, {Gai}, {Garcia-Reinaldos}, {Guerra}, {Hambly}, {Hauser},
  {Jan{\ss}en}, {Jordan}, {Kostrzewa-Rutkowska}, {Lattanzi}, {Liao}, {Licata},
  {Lister}, {L{\"o}ffler}, {Marchant}, {Masip}, {Mignard}, {Mints}, {Molina},
  {Mora}, {Morbidelli}, {Murphy}, {Pagani}, {Panuzzo}, {Pe{\~n}alosa Esteller},
  {Poggio}, {Re Fiorentin}, {Riva}, {Sagrist{\`a} Sell{\'e}s}, {Sanchez
  Gimenez}, {Sarasso}, {Sciacca}, {Siddiqui}, {Smart}, {Souami}, {Spagna},
  {Steele}, {Taris}, {Utrilla}, {van Reeven}, \& {Vecchiato}}]{Lindegren2021}
{Lindegren}, L., {Klioner}, S.~A., {Hern{\'a}ndez}, J., {et~al.} 2021,
  \bibinfo{title}{{Gaia Early Data Release 3. The astrometric solution},} \aap,
  649, A2, \dodoi{10.1051/0004-6361/202039709}

\bibitem[{J.~J. {Lissauer} {et~al.}(2023){Lissauer}, {Batalha}, \&
  {Borucki}}]{Lissauer2023}
{Lissauer}, J.~J., {Batalha}, N.~M., \& {Borucki}, W.~J. 2023, in Astronomical
  Society of the Pacific Conference Series, Vol. 534, Protostars and Planets
  VII, ed. S.~{Inutsuka}, Y.~{Aikawa}, T.~{Muto}, K.~{Tomida}, \& M.~{Tamura},
  839, \dodoi{10.48550/arXiv.2311.04981}

\bibitem[{T. {Mazeh} {et~al.}(2016){Mazeh}, {Holczer}, \&
  {Faigler}}]{Mazeh2016}
{Mazeh}, T., {Holczer}, T., \& {Faigler}, S. 2016, \bibinfo{title}{{Dearth of
  short-period Neptunian exoplanets: A desert in period-mass and period-radius
  planes},} \aap, 589, A75, \dodoi{10.1051/0004-6361/201528065}

\bibitem[{F. {Ochsenbein} {et~al.}(2000){Ochsenbein}, {Bauer}, \&
  {Marcout}}]{vizier}
{Ochsenbein}, F., {Bauer}, P., \& {Marcout}, J. 2000, \bibinfo{title}{{The
  VizieR database of astronomical catalogues},} \aaps, 143, 23,
  \dodoi{10.1051/aas:2000169}

\bibitem[{S.~S.~R. {Offner} {et~al.}(2023){Offner}, {Moe}, {Kratter},
  {Sadavoy}, {Jensen}, \& {Tobin}}]{Offner2023}
{Offner}, S.~S.~R., {Moe}, M., {Kratter}, K.~M., {et~al.} 2023, in Astronomical
  Society of the Pacific Conference Series, Vol. 534, Protostars and Planets
  VII, ed. S.~{Inutsuka}, Y.~{Aikawa}, T.~{Muto}, K.~{Tomida}, \& M.~{Tamura},
  275, \dodoi{10.48550/arXiv.2203.10066}

\bibitem[{B. {Paxton} {et~al.}(2011){Paxton}, {Bildsten}, {Dotter}, {Herwig},
  {Lesaffre}, \& {Timmes}}]{Paxton2011}
{Paxton}, B., {Bildsten}, L., {Dotter}, A., {et~al.} 2011,
  \bibinfo{title}{{Modules for Experiments in Stellar Astrophysics (MESA)},}
  \apjs, 192, 3, \dodoi{10.1088/0067-0049/192/1/3}

\bibitem[{B. {Paxton} {et~al.}(2013){Paxton}, {Cantiello}, {Arras}, {Bildsten},
  {Brown}, {Dotter}, {Mankovich}, {Montgomery}, {Stello}, {Timmes}, \&
  {Townsend}}]{Paxton2013}
{Paxton}, B., {Cantiello}, M., {Arras}, P., {et~al.} 2013,
  \bibinfo{title}{{Modules for Experiments in Stellar Astrophysics (MESA):
  Planets, Oscillations, Rotation, and Massive Stars},} \apjs, 208, 4,
  \dodoi{10.1088/0067-0049/208/1/4}

\bibitem[{B. {Paxton} {et~al.}(2015){Paxton}, {Marchant}, {Schwab}, {Bauer},
  {Bildsten}, {Cantiello}, {Dessart}, {Farmer}, {Hu}, {Langer}, {Townsend},
  {Townsley}, \& {Timmes}}]{Paxton2015}
{Paxton}, B., {Marchant}, P., {Schwab}, J., {et~al.} 2015,
  \bibinfo{title}{{Modules for Experiments in Stellar Astrophysics (MESA):
  Binaries, Pulsations, and Explosions},} \apjs, 220, 15,
  \dodoi{10.1088/0067-0049/220/1/15}

\bibitem[{F. Pedregosa {et~al.}(2011)Pedregosa, Varoquaux, Gramfort, Michel,
  Thirion, Grisel, Blondel, Prettenhofer, Weiss, Dubourg, Vanderplas, Passos,
  Cournapeau, Brucher, Perrot, \& Duchesnay}]{scikit-learn}
Pedregosa, F., Varoquaux, G., Gramfort, A., {et~al.} 2011,
  \bibinfo{title}{Scikit-learn: Machine Learning in {P}ython,} Journal of
  Machine Learning Research, 12, 2825

\bibitem[{E.~A. {Petigura} {et~al.}(2013){Petigura}, {Howard}, \&
  {Marcy}}]{Petigura2013}
{Petigura}, E.~A., {Howard}, A.~W., \& {Marcy}, G.~W. 2013,
  \bibinfo{title}{{Prevalence of Earth-size planets orbiting Sun-like stars},}
  Proceedings of the National Academy of Science, 110, 19273,
  \dodoi{10.1073/pnas.1319909110}

\bibitem[{E.~A. {Petigura} {et~al.}(2022){Petigura}, {Rogers}, {Isaacson},
  {Owen}, {Kraus}, {Winn}, {MacDougall}, {Howard}, {Fulton}, {Kosiarek},
  {Weiss}, {Behmard}, \& {Blunt}}]{Petigura2022}
{Petigura}, E.~A., {Rogers}, J.~G., {Isaacson}, H., {et~al.} 2022,
  \bibinfo{title}{{The California-Kepler Survey. X. The Radius Gap as a
  Function of Stellar Mass, Metallicity, and Age},} \aj, 163, 179,
  \dodoi{10.3847/1538-3881/ac51e3}

\bibitem[{D. {Raghavan} {et~al.}(2010){Raghavan}, {McAlister}, {Henry},
  {Latham}, {Marcy}, {Mason}, {Gies}, {White}, \& {ten
  Brummelaar}}]{Raghavan2010}
{Raghavan}, D., {McAlister}, H.~A., {Henry}, T.~J., {et~al.} 2010,
  \bibinfo{title}{{A Survey of Stellar Families: Multiplicity of Solar-type
  Stars},} \apjs, 190, 1, \dodoi{10.1088/0067-0049/190/1/1}

\bibitem[{H. {Rauer} {et~al.}(2014){Rauer}, {Catala}, {Aerts}, {Appourchaux},
  {Benz}, {Brandeker}, {Christensen-Dalsgaard}, {Deleuil}, {Gizon}, {Goupil},
  {G{\"u}del}, {Janot-Pacheco}, {Mas-Hesse}, {Pagano}, {Piotto}, {Pollacco},
  {Santos}, {Smith}, {Su{\'a}rez}, {Szab{\'o}}, {Udry}, {Adibekyan}, {Alibert},
  {Almenara}, {Amaro-Seoane}, {Eiff}, {Asplund}, {Antonello}, {Barnes},
  {Baudin}, {Belkacem}, {Bergemann}, {Bihain}, {Birch}, {Bonfils}, {Boisse},
  {Bonomo}, {Borsa}, {Brand{\~a}o}, {Brocato}, {Brun}, {Burleigh}, {Burston},
  {Cabrera}, {Cassisi}, {Chaplin}, {Charpinet}, {Chiappini}, {Church},
  {Csizmadia}, {Cunha}, {Damasso}, {Davies}, {Deeg}, {D{\'\i}az}, {Dreizler},
  {Dreyer}, {Eggenberger}, {Ehrenreich}, {Eigm{\"u}ller}, {Erikson}, {Farmer},
  {Feltzing}, {de Oliveira Fialho}, {Figueira}, {Forveille}, {Fridlund},
  {Garc{\'\i}a}, {Giommi}, {Giuffrida}, {Godolt}, {Gomes da Silva}, {Granzer},
  {Grenfell}, {Grotsch-Noels}, {G{\"u}nther}, {Haswell}, {Hatzes},
  {H{\'e}brard}, {Hekker}, {Helled}, {Heng}, {Jenkins}, {Johansen},
  {Khodachenko}, {Kislyakova}, {Kley}, {Kolb}, {Krivova}, {Kupka}, {Lammer},
  {Lanza}, {Lebreton}, {Magrin}, {Marcos-Arenal}, {Marrese}, {Marques},
  {Martins}, {Mathis}, {Mathur}, {Messina}, {Miglio}, {Montalban}, {Montalto},
  {Monteiro}, {Moradi}, {Moravveji}, {Mordasini}, {Morel}, {Mortier},
  {Nascimbeni}, {Nelson}, {Nielsen}, {Noack}, {Norton}, {Ofir}, {Oshagh},
  {Ouazzani}, {P{\'a}pics}, {Parro}, {Petit}, {Plez}, {Poretti}, {Quirrenbach},
  {Ragazzoni}, {Raimondo}, {Rainer}, {Reese}, {Redmer}, {Reffert},
  {Rojas-Ayala}, {Roxburgh}, {Salmon}, {Santerne}, {Schneider}, {Schou},
  {Schuh}, {Schunker}, {Silva-Valio}, {Silvotti}, {Skillen}, {Snellen}, {Sohl},
  {Sousa}, {Sozzetti}, {Stello}, {Strassmeier}, {{\v{S}}vanda}, {Szab{\'o}},
  {Tkachenko}, {Valencia}, {Van Grootel}, {Vauclair}, {Ventura}, {Wagner},
  {Walton}, {Weingrill}, {Werner}, {Wheatley}, \& {Zwintz}}]{Rauer2014}
{Rauer}, H., {Catala}, C., {Aerts}, C., {et~al.} 2014, \bibinfo{title}{{The
  PLATO 2.0 mission},} Experimental Astronomy, 38, 249,
  \dodoi{10.1007/s10686-014-9383-4}

\bibitem[{G.~R. {Ricker} {et~al.}(2015){Ricker}, {Winn}, {Vanderspek},
  {Latham}, {Bakos}, {Bean}, {Berta-Thompson}, {Brown}, {Buchhave}, {Butler},
  {Butler}, {Chaplin}, {Charbonneau}, {Christensen-Dalsgaard}, {Clampin},
  {Deming}, {Doty}, {De Lee}, {Dressing}, {Dunham}, {Endl}, {Fressin}, {Ge},
  {Henning}, {Holman}, {Howard}, {Ida}, {Jenkins}, {Jernigan}, {Johnson},
  {Kaltenegger}, {Kawai}, {Kjeldsen}, {Laughlin}, {Levine}, {Lin}, {Lissauer},
  {MacQueen}, {Marcy}, {McCullough}, {Morton}, {Narita}, {Paegert}, {Palle},
  {Pepe}, {Pepper}, {Quirrenbach}, {Rinehart}, {Sasselov}, {Sato}, {Seager},
  {Sozzetti}, {Stassun}, {Sullivan}, {Szentgyorgyi}, {Torres}, {Udry}, \&
  {Villasenor}}]{Ricker2015}
{Ricker}, G.~R., {Winn}, J.~N., {Vanderspek}, R., {et~al.} 2015,
  \bibinfo{title}{{Transiting Exoplanet Survey Satellite (TESS)},} Journal of
  Astronomical Telescopes, Instruments, and Systems, 1, 014003,
  \dodoi{10.1117/1.JATIS.1.1.014003}

\bibitem[{S. {Shibata} \& A. {Izidoro}(2025){Shibata} \&
  {Izidoro}}]{Shibata2025}
{Shibata}, S., \& {Izidoro}, A. 2025, \bibinfo{title}{{A tale of dynamical
  instabilities and giant impacts in the radius valley},} arXiv e-prints,
  arXiv:2505.23943, \dodoi{10.48550/arXiv.2505.23943}

\bibitem[{K. {Sullivan} {et~al.}(2025){Sullivan}, {Kraus}, {Berger}, \&
  {Huber}}]{Sullivan2025a}
{Sullivan}, K., {Kraus}, A.~L., {Berger}, T.~A., \& {Huber}, D. 2025,
  \bibinfo{title}{{Quantifying the Contamination from nearby Stellar Companions
  in Gaia DR3 Photometry},} \aj, 169, 29, \dodoi{10.3847/1538-3881/ad9330}

\bibitem[{K. {Sullivan} {et~al.}(2022){Sullivan}, {Kraus}, \&
  {Mann}}]{Sullivan2022b}
{Sullivan}, K., {Kraus}, A.~L., \& {Mann}, A.~W. 2022,
  \bibinfo{title}{{Revising Properties of Planet-Host Binary Systems. I.
  Methods and Pilot Study},} \apj, 935, 141, \dodoi{10.3847/1538-4357/ac7be9}

\bibitem[{K. {Sullivan} {et~al.}(2023){Sullivan}, {Kraus}, {Huber}, {Petigura},
  {Evans}, {Dupuy}, {Zhang}, {Berger}, {Gaidos}, \& {Mann}}]{Sullivan2023}
{Sullivan}, K., {Kraus}, A.~L., {Huber}, D., {et~al.} 2023,
  \bibinfo{title}{{Revising Properties of Planet-Host Binary Systems. III.
  There Is No Observed Radius Gap for Kepler Planets in Binary Star Systems},}
  \aj, 165, 177, \dodoi{10.3847/1538-3881/acbdf9}

\bibitem[{K. {Sullivan} {et~al.}(2024){Sullivan}, {Kraus}, {Berger}, {Dupuy},
  {Evans}, {Gaidos}, {Huber}, {Ireland}, {Mann}, {Petigura}, {Thao}, {Wood}, \&
  {Zhang}}]{Sullivan2024b}
{Sullivan}, K., {Kraus}, A.~L., {Berger}, T.~A., {et~al.} 2024,
  \bibinfo{title}{{Revising Properties of Planet{\textendash}Host Binary
  Systems. IV. The Radius Distribution of Small Planets in Binary Star Systems
  Is Dependent on Stellar Separation},} \aj, 168, 129,
  \dodoi{10.3847/1538-3881/ad6310}

\bibitem[{S.~E. {Thompson} {et~al.}(2018){Thompson}, {Coughlin}, {Hoffman},
  {Mullally}, {Christiansen}, {Burke}, {Bryson}, {Batalha}, {Haas},
  {Catanzarite}, {Rowe}, {Barentsen}, {Caldwell}, {Clarke}, {Jenkins}, {Li},
  {Latham}, {Lissauer}, {Mathur}, {Morris}, {Seader}, {Smith}, {Klaus},
  {Twicken}, {Van Cleve}, {Wohler}, {Akeson}, {Ciardi}, {Cochran}, {Henze},
  {Howell}, {Huber}, {Pr{\v{s}}a}, {Ram{\'\i}rez}, {Morton}, {Barclay},
  {Campbell}, {Chaplin}, {Charbonneau}, {Christensen-Dalsgaard}, {Dotson},
  {Doyle}, {Dunham}, {Dupree}, {Ford}, {Geary}, {Girouard}, {Isaacson},
  {Kjeldsen}, {Quintana}, {Ragozzine}, {Shabram}, {Shporer}, {Silva Aguirre},
  {Steffen}, {Still}, {Tenenbaum}, {Welsh}, {Wolfgang}, {Zamudio}, {Koch}, \&
  {Borucki}}]{Thompson2018}
{Thompson}, S.~E., {Coughlin}, J.~L., {Hoffman}, K., {et~al.} 2018,
  \bibinfo{title}{{Planetary Candidates Observed by Kepler. VIII. A Fully
  Automated Catalog with Measured Completeness and Reliability Based on Data
  Release 25},} \apjs, 235, 38, \dodoi{10.3847/1538-4365/aab4f9}

\bibitem[{V. {Van Eylen} {et~al.}(2018){Van Eylen}, {Agentoft}, {Lundkvist},
  {Kjeldsen}, {Owen}, {Fulton}, {Petigura}, \& {Snellen}}]{VanEylen2018}
{Van Eylen}, V., {Agentoft}, C., {Lundkvist}, M.~S., {et~al.} 2018,
  \bibinfo{title}{{An asteroseismic view of the radius valley: stripped cores,
  not born rocky},} \mnras, 479, 4786, \dodoi{10.1093/mnras/sty1783}

\bibitem[{P. Virtanen {et~al.}(2020)Virtanen, Gommers, Oliphant, Haberland,
  Reddy, Cournapeau, Burovski, Peterson, Weckesser, Bright, {van der Walt},
  Brett, Wilson, Millman, Mayorov, Nelson, Jones, Kern, Larson, Carey, Polat,
  Feng, Moore, {VanderPlas}, Laxalde, Perktold, Cimrman, Henriksen, Quintero,
  Harris, Archibald, Ribeiro, Pedregosa, {van Mulbregt}, \& {SciPy 1.0
  Contributors}}]{Virtanen2020}
Virtanen, P., Gommers, R., Oliphant, T.~E., {et~al.} 2020,
  \bibinfo{title}{{{SciPy} 1.0: Fundamental Algorithms for Scientific Computing
  in Python},} Nature Methods, 17, 261, \dodoi{10.1038/s41592-019-0686-2}

\bibitem[{J. {Wang} {et~al.}(2014){Wang}, {Fischer}, {Xie}, \&
  {Ciardi}}]{Wang2014}
{Wang}, J., {Fischer}, D.~A., {Xie}, J.-W., \& {Ciardi}, D.~R. 2014,
  \bibinfo{title}{{Influence of Stellar Multiplicity on Planet Formation. II.
  Planets are Less Common in Multiple-star Systems with Separations Smaller
  than 1500 AU},} \apj, 791, 111, \dodoi{10.1088/0004-637X/791/2/111}

\bibitem[{J. {Wang} {et~al.}(2015){Wang}, {Fischer}, {Xie}, \&
  {Ciardi}}]{Wang2015}
{Wang}, J., {Fischer}, D.~A., {Xie}, J.-W., \& {Ciardi}, D.~R. 2015,
  \bibinfo{title}{{Influence of Stellar Multiplicity on Planet Formation. IV.
  Adaptive Optics Imaging of Kepler Stars with Multiple Transiting Planet
  Candidates},} \apj, 813, 130, \dodoi{10.1088/0004-637X/813/2/130}

\bibitem[{J.~G. {Winters} {et~al.}(2019){Winters}, {Henry}, {Jao},
  {Subasavage}, {Chatelain}, {Slatten}, {Riedel}, {Silverstein}, \&
  {Payne}}]{Winters2019}
{Winters}, J.~G., {Henry}, T.~J., {Jao}, W.-C., {et~al.} 2019,
  \bibinfo{title}{{The Solar Neighborhood. XLV. The Stellar Multiplicity Rate
  of M Dwarfs Within 25 pc},} \aj, 157, 216, \dodoi{10.3847/1538-3881/ab05dc}

\bibitem[{M.~L. {Wood} {et~al.}(2021){Wood}, {Mann}, \& {Kraus}}]{Wood2021}
{Wood}, M.~L., {Mann}, A.~W., \& {Kraus}, A.~L. 2021,
  \bibinfo{title}{{Characterizing Undetected Stellar Companions with Combined
  Data Sets},} \aj, 162, 128, \dodoi{10.3847/1538-3881/ac0ae9}

\bibitem[{G. {Zhou} {et~al.}(2019){Zhou}, {Huang}, {Bakos}, {Hartman},
  {Latham}, {Quinn}, {Collins}, {Winn}, {Wong}, {Kov{\'a}cs}, {Csubry},
  {Bhatti}, {Penev}, {Bieryla}, {Esquerdo}, {Berlind}, {Calkins}, {de
  Val-Borro}, {Noyes}, {L{\'a}z{\'a}r}, {Papp}, {S{\'a}ri}, {Kov{\'a}cs},
  {Buchhave}, {Szklenar}, {B{\'e}ky}, {Johnson}, {Cochran}, {Kniazev},
  {Stassun}, {Fulton}, {Shporer}, {Espinoza}, {Bayliss}, {Everett}, {Howell},
  {Hellier}, {Anderson}, {Collier Cameron}, {West}, {Brown}, {Schanche},
  {Barkaoui}, {Pozuelos}, {Gillon}, {Jehin}, {Benkhaldoun}, {Daassou},
  {Ricker}, {Vanderspek}, {Seager}, {Jenkins}, {Lissauer}, {Armstrong},
  {Collins}, {Gan}, {Hart}, {Horne}, {Kielkopf}, {Nielsen}, {Nishiumi},
  {Narita}, {Palle}, {Relles}, {Sefako}, {Tan}, {Davies}, {Goeke}, {Guerrero},
  {Haworth}, \& {Villanueva}}]{Zhou2019}
{Zhou}, G., {Huang}, C.~X., {Bakos}, G.~{\'A}., {et~al.} 2019,
  \bibinfo{title}{{Two New HATNet Hot Jupiters around A Stars and the First
  Glimpse at the Occurrence Rate of Hot Jupiters from TESS},} \aj, 158, 141,
  \dodoi{10.3847/1538-3881/ab36b5}

\bibitem[{C. {Ziegler} {et~al.}(2020){Ziegler}, {Tokovinin}, {Brice{\~n}o},
  {Mang}, {Law}, \& {Mann}}]{Ziegler2020}
{Ziegler}, C., {Tokovinin}, A., {Brice{\~n}o}, C., {et~al.} 2020,
  \bibinfo{title}{{SOAR TESS Survey. I. Sculpting of TESS Planetary Systems by
  Stellar Companions},} \aj, 159, 19, \dodoi{10.3847/1538-3881/ab55e9}

\bibitem[{C. {Ziegler} {et~al.}(2017){Ziegler}, {Law}, {Morton}, {Baranec},
  {Riddle}, {Atkinson}, {Baker}, {Roberts}, \& {Ciardi}}]{Ziegler2017}
{Ziegler}, C., {Law}, N.~M., {Morton}, T., {et~al.} 2017,
  \bibinfo{title}{{Robo-AO Kepler Planetary Candidate Survey. III. Adaptive
  Optics Imaging of 1629 Kepler Exoplanet Candidate Host Stars},} \aj, 153, 66,
  \dodoi{10.3847/1538-3881/153/2/66}

\bibitem[{C. {Ziegler} {et~al.}(2018){Ziegler}, {Law}, {Baranec}, {Riddle},
  {Duev}, {Howard}, {Jensen-Clem}, {Kulkarni}, {Morton}, \&
  {Salama}}]{Ziegler2018}
{Ziegler}, C., {Law}, N.~M., {Baranec}, C., {et~al.} 2018,
  \bibinfo{title}{{Robo-AO Kepler Survey. IV. The Effect of Nearby Stars on
  3857 Planetary Candidate Systems},} \aj, 155, 161,
  \dodoi{10.3847/1538-3881/aab042}

\end{thebibliography}

\begin{longrotatetable}
\begin{deluxetable*}{ccCCCCCCCCCC}
\setcounter{table}{0}
    \tablecaption{Distributions of Derived Stellar Properties for the Probabilistic Binary Catalog \label{tab:stars}}
    \tabletypesize{\small}
\tablecolumns{12}
\tablehead{\colhead{Gaia DR2} & \colhead{Kepler ID} & \colhead{Distance}     & \colhead{Separation}     & \colhead{M$_{\rm pri}$}     & \colhead{Mass ratio}     & \colhead{$T_{\rm eff, pri}$}     & \colhead{$L_{\rm pri}$}    & \colhead{$T_{\rm eff, sec}$}    & \colhead{$L_{\rm sec}$}    & \colhead{PRCF$_{\rm pri}$}    & \colhead{PRCF$_{\rm sec}$}\\    \colhead{} & \colhead{} & \colhead{pc}     & \colhead{au}     & \colhead{M$_{\odot}$}     & \colhead{}     & \colhead{K}     & \colhead{$L_{\odot}$}    & \colhead{K}    & \colhead{$L_{\odot}$}    & \colhead{}    & \colhead{}}
\startdata
2049134982911348480 & 3459113 & 607 & 11$^{+37}_{-8}$ & 1.07$^{+0.07}_{-0.07}$ &        0.76$^{+0.34}_{-0.29}$ & 5911$^{+112}_{-109}$ & 1.50$^{+0.10}_{-0.11}$ &        3797$^{+1293}_{-541}$ & 0.42$^{+0.19}_{-0.21}$ & 1.01$^{+0.14}_{-0.01}$ & 6.90$^{+17.61}_{-4.87}$\\ 
2049135390918975488 & 3459226 & 299 & 10$^{+22}_{-7}$ & 1.56$^{+0.03}_{-0.04}$ &        0.26$^{+0.33}_{-0.32}$ & 6828$^{+38}_{-77}$ & 6.23$^{+0.53}_{-0.52}$ &        4724$^{+1748}_{-1274}$ & 0.51$^{+0.65}_{-0.27}$ & 1.00$^{+0.12}_{-0.00}$ & 13.50$^{+35.84}_{-11.33}$\\ 
2049136082423106560 & 3459775 & 1205 & 19$^{+70}_{-16}$ & 1.44$^{+0.07}_{-0.07}$ &        0.46$^{+0.33}_{-0.32}$ & 6497$^{+128}_{-131}$ & 5.83$^{+0.50}_{-0.51}$ &        4531$^{+1815}_{-1116}$ & 3.11$^{+0.96}_{-2.32}$ & 1.01$^{+0.14}_{-0.01}$ & 7.25$^{+34.72}_{-5.20}$\\ 
2049136391660806656 & 3459976 & 1631 & 20$^{+76}_{-17}$ & 1.06$^{+0.08}_{-0.08}$ &        0.53$^{+0.35}_{-0.32}$ & 5887$^{+115}_{-110}$ & 1.80$^{+0.32}_{-0.31}$ &        3874$^{+1462}_{-538}$ & 0.75$^{+0.36}_{-0.36}$ & 1.01$^{+0.18}_{-0.01}$ & 5.96$^{+13.16}_{-4.13}$\\ 
2049136632178988544 & 3560803 & 723 & 13$^{+42}_{-10}$ & 0.77$^{+0.05}_{-0.05}$ &        0.41$^{+0.36}_{-0.30}$ & 5120$^{+98}_{-98}$ & 0.34$^{+0.04}_{-0.04}$ &        3586$^{+627}_{-361}$ & 0.03$^{+0.03}_{-0.02}$ & 1.01$^{+0.09}_{-0.01}$ & 5.98$^{+9.17}_{-3.61}$\\ 
2049137319373726720 & 3560585 & 849 & 13$^{+43}_{-10}$ & 0.87$^{+0.03}_{-0.02}$ &        0.29$^{+0.36}_{-0.29}$ & 5249$^{+73}_{-70}$ & 0.89$^{+0.07}_{-0.07}$ &        3625$^{+882}_{-396}$ & 0.45$^{+0.08}_{-0.08}$ & 1.01$^{+0.13}_{-0.01}$ & 6.02$^{+10.27}_{-3.97}$\\ 
2050192197690663424 & 1028012 & 1100 & 18$^{+63}_{-15}$ & 1.26$^{+0.09}_{-0.09}$ &        0.01$^{+0.34}_{-0.33}$ & 6097$^{+112}_{-110}$ & 3.61$^{+0.29}_{-0.29}$ &        4211$^{+1679}_{-879}$ & 2.13$^{+0.38}_{-0.41}$ & 1.02$^{+0.29}_{-0.02}$ & 4.74$^{+18.40}_{-3.19}$\\ 
2050233601176543232 & 757099 & 369 & 9$^{+23}_{-7}$ & 0.87$^{+0.05}_{-0.05}$ &        0.40$^{+0.33}_{-0.32}$ & 5364$^{+95}_{-93}$ & 0.86$^{+0.05}_{-0.05}$ &        3668$^{+828}_{-401}$ & 0.36$^{+0.07}_{-0.08}$ & 1.01$^{+0.11}_{-0.01}$ & 5.94$^{+10.03}_{-3.75}$\\ 
2050236143797351936 & 1026474 & 370 & 7$^{+17}_{-5}$ & 0.65$^{+0.02}_{-0.02}$ &        0.93$^{+0.37}_{-0.29}$ & 4177$^{+64}_{-63}$ & 0.12$^{+0.01}_{-0.01}$ &        3474$^{+389}_{-314}$ & 0.02$^{+0.01}_{-0.01}$ & 1.04$^{+0.16}_{-0.03}$ & 3.65$^{+4.70}_{-1.84}$\\ 
2050239511051851264 & 1027016 & 458 & 10$^{+31}_{-8}$ & 0.82$^{+0.02}_{-0.02}$ &        0.10$^{+0.34}_{-0.31}$ & 5058$^{+72}_{-71}$ & 0.50$^{+0.03}_{-0.03}$ &        3634$^{+737}_{-387}$ & 0.16$^{+0.04}_{-0.04}$ & 1.02$^{+0.13}_{-0.02}$ & 5.15$^{+8.48}_{-3.11}$\\ 
\enddata
\tablecomments{The first 10 rows of the stellar catalog. The remainder of the catalog is available online in machine-readable format.}
\end{deluxetable*}
\end{longrotatetable}

\end{document}